\documentclass[11pt,letterpaper]{article} 
\pdfoutput=1
\usepackage{amsmath,amssymb,amsfonts}
\usepackage[usenames,dvipsnames]{xcolor}
\definecolor{airforceblue}{rgb}{0.36, 0.54, 0.66}
\definecolor{meb}{rgb}{0.01, 0.31, 0.59}
\usepackage[T1]{fontenc}
\usepackage{palatino}
\def\title{
Null Raychaudhuri:\\ Canonical Structure and the  Dressing Time} 
\usepackage{graphicx}
\usepackage[colorlinks=true, pdfstartview=FitV, linkcolor=blue, citecolor=magenta, urlcolor=blue]{hyperref}
\usepackage{cancel}
\usepackage{mdframed,framed}
\usepackage[usenames,dvipsnames]{xcolor}
\usepackage{mathpazo}
\usepackage[normalem]{ulem}
\usepackage{layout}
\usepackage[nosort]{cite}

\newcommand{\D}{\text{d}}
\newcommand{\beq}{\begin{eqnarray}}
\newcommand{\eeq}{\end{eqnarray}}
\newcommand{\beqn}{\begin{eqnarray}}
\newcommand{\eeqn}{\end{eqnarray}}
\newcommand{\pa}{\partial}

\newcommand{\cL}{{\cal L}}

\newcommand{\fL}{\mathfrak{L}}

\usepackage{tikz-cd}
\usepackage{pict2e}
\makeatletter
\newcommand{\variable@rule}[1]{%
  \fontdimen8  
  \ifx#1\displaystyle\textfont3\else
    \ifx#1\textstyle\textfont3\else
      \ifx#1\scriptstyle\scriptfont3\else
        \scriptscriptfont3\relax
  \fi\fi\fi
}
\usepackage{mathtools}

\usepackage{tocloft}
\newcommand{\ve}{\varepsilon}

\newcommand{\cN}{\cal{N}}
\newcommand{\cC}{{\cal{C}}}
\newcommand{\cF}{{\cal{F}}}
\newcommand{\cP}{{\cal{{P}}}}
\newcommand{\bcP}{{\overline{\cal{P}}}}
\newcommand{\bq}{{\overline{q}}}

\newcommand{\bz}{{\overline{z}}}
\newcommand{\oD}{{\overline{D}}}
\newcommand{\bzeta}{{\overline{\zeta}}}
\newcommand{\Bm}{{\overline{m}}}
\newcommand{\bD}{{\overline{\Delta}}}
\newcommand{\rd}{\text{d}}

\newcommand{\chkM}{{\color{red} \,\checkmark\kern-5pt{}_{M}}}

\newcommand{\be}{\begin{equation}}
\newcommand{\ee}{\end{equation}}
\newcommand{\bea}{\begin{eqnarray}}
\newcommand{\eea}{\end{eqnarray}}

\def\pa{\partial}

\newcommand{\ctime}{dressing time }

\newcommand{\perimeter}[1]{
	\centerline{
		\begin{minipage}[c]{0.7\textwidth}
			\begin{center}
			$^a$ Perimeter Institute for Theoretical Physics,\\
			 31 Caroline St. N., Waterloo ON, Canada, N2L 2Y5
			\end{center}
		\end{minipage}
		}
	}
\newcommand{\uiuc}[1]{
	\centerline{
		\begin{minipage}[c]{0.8\textwidth}
			\begin{center}
		$^b$ Illinois Center for Advanced Studies of the Universe \& Department of Physics,\\
			University of Illinois, 1110 West Green St., Urbana IL 61801, U.S.A.
			\end{center}
		\end{minipage}
		}
	}

\begin{document}

{\centering
 \vspace*{1cm}
\textbf{\LARGE{\title{}}}
\vspace{0.5cm}
\begin{center}
Luca Ciambelli,$^a$ Laurent Freidel,$^a$ Robert G. Leigh$^{b,a}$\\
\vspace{0.5cm}
\textit{\perimeter{}}\\
\vspace{0.5cm}
\textit{\uiuc{}}
\end{center}
\vspace{0.5cm}
{\small{\href{mailto:ciambelli.luca@gmail.com}{ciambelli.luca@gmail.com}, \ \href{mailto:lfreidel@perimeterinstitute.ca}{lfreidel@perimeterinstitute.ca}}}, \  \href{mailto:rgleigh@illinois.edu}{rgleigh@illinois.edu}
\vspace{1cm}
\begin{abstract}
\vspace{0.5cm}
We initiate a study of gravity focusing on generic null hypersurfaces, non-perturbatively in the Newton coupling. 
We present an off-shell account of the extended phase space of the theory, which includes the expected spin-2 data as well as spin-0, spin-1 and arbitrary matter degrees of freedom. We construct the charges and the corresponding kinematic Poisson brackets, employing a Beltrami parameterization of the spin-2 modes. We explicitly show that the constraint algebra closes, the details of which depend on the non-perturbative mixing between spin-0 and spin-2 modes. Finally we show that the spin zero sector encodes a notion of a clock, called dressing time, which is dynamical and conjugate to the constraint.

It is well-known that the null Raychaudhuri equation describes how the geometric data of a null hypersurface evolve in null time in response to gravitational radiation and external matter.  Our analysis leads to three complementary viewpoints on this equation. First, it can be understood as a Carrollian stress tensor conservation equation. Second, we construct spin-$0$, spin-$2$ and matter stress tensors that act as generators of null time reparametrizations for each sector.
This leads to the perspective that the null Raychaudhuri equation can be understood as imposing  that the sum of CFT-like stress tensors vanishes.
Third, we  solve the Raychaudhuri constraint non-perturbatively. The solution relates the dressing time to the spin-$2$ and matter boost charge operators.

Finally we establish that the corner charge corresponding to the boost operator in the dressing time frame is monotonic.  These results show that the notion of an observer can be thought of as emerging from the gravitational degrees of freedom themselves. We briefly mention that the construction offers new insights into focusing conjectures. 
\end{abstract}}

\thispagestyle{empty}

\newpage
\tableofcontents
\thispagestyle{empty}
\newpage
\clearpage
\pagenumbering{arabic} 


\section{Introduction}

It is a familiar property of classical gravitational theories such as general relativity that diffeomorphisms in spacetime play the role of a gauge symmetry. Working on a fixed background such as Minkowski spacetime, when suitable gauge fixing and linearization is performed, the field equations have wave solutions. The simplest notion of quantization then gives rise to corresponding helicity-2 graviton particle states, which can be regarded as the most basic prediction of a quantum theory. Critically, it is in this sense in which quantum gravity may have a correspondence limit, reducing to classical gravity at least in the weak-field regime. 
This is however quite far from a full quantum gravitational theory, in which one should expect background independence along with a formulation that does not rely on a corresponding perturbative expansion in the Newton coupling. In fact, it is well-known that if one approaches quantum gravity from the latter point of view in terms of an effective Lagrangian field theory, then that theory is non-renormalizable, signaling that new physics is required at high energy and/or short distance scales. It is also not obvious that there is not new physics in the infrared as well, which may be associated with complications that may occur in the correspondence limit.

There are several top-down approaches to quantum gravity postulating new physics at high energy scales. The new physics either comes under the form of an infinite tower of new  degrees of freedom as in string theory or under some fundamental postulate of discreteness of  geometry as in loop quantum gravity or causal dynamical triangulations.
The perspective that we plan to pursue  here  is more modest and aims at a bottom-up approach towards quantum geometry relying on a precise and non-perturbative semi-classical understanding of the coupling between the matter and gravitational degree of freedom and the backreaction it implies on null geometry. The goal is to provide some basic unifying features that ought to be present in the semi-classical limit of any top-down approach to quantum gravity. 

Top-down approaches to quantum gravity have in common one feature, that performing a semi-classical limit is not necessarily straightforward. This is mainly due to the fact that a theory of quantum gravity needs to be fundamentally non-local and this in turn implies that it is challenging to match at the quantum level  the classical symmetries.  Indeed, classical gravity possesses many symmetries: some are trivial redundancies, while some are physical symmetries that organize observables. This is an old discussion, going back to the original works of Noether \cite{Noether1918}. The point here is that while diffeomorphisms are gauge symmetries and thus the theory must be discussed, either classically or quantum mechanically, with the corresponding constraints, the charge operators do not vanish even on-shell, but localize on codimension-2 surfaces (which herein will be referred to as 'corners' for brevity). It has been proposed in recent years that corners play a central role in any gravitational theory, an idea that has been referred to as the corner proposal \cite{Freidel:2015gpa, Donnelly:2016auv, Geiller:2017xad, Speranza:2017gxd, Geiller:2017whh, Donnelly:2020xgu, Ciambelli:2021vnn, Freidel:2021cjp, Ciambelli:2022cfr, Donnelly:2022kfs} (see also the reviews \cite{Ciambelli:2022vot, Freidel:2023bnj, Ciambelli:2023bmn} and references therein). 
Universal algebras of charges at corners have been found, while their quantum counterparts remain to be fully understood.

Where are these symmetries and how are the observables organized in classical and then quantum gravity? A bottom-up approach to quantum gravity, while potentially incomplete, has the virtue of automatically addressing these questions. Indeed, in this approach classical gravity is not the endpoint of a limiting procedure, it is the starting point toward the quantum regime, in which symmetries can be incorporated from the start.
Whether one intends to address classical evolution or to quantize the theory, a choice of Cauchy surface is made, which given a classical spacetime would be a hypersurface that is either timelike or null. Once chosen, there are several notions of phase space that we might address.  First, there is the phase space arising from some choice of fields in spacetime (leading to fields on the hypersurface). In this case generally, because of the presence of gauge symmetries, the symplectic structure derived from a Lagrangian would be degenerate (i.e., it is presymplectic). Second there is a reduced phase space where one eliminates some subset of the fields as 'pure gauge,' thus obtaining perhaps a simpler theory with a non-degenerate symplectic structure.  In each case, there is a further complication in that the (gauge) symmetries may or may not be canonically represented on the phase space. In this regard, it has recently been realized \cite{Ciambelli:2021nmv,Ciambelli:2021vnn, Freidel:2021dxw,Klinger:2023qna}
 that one may first pass to an extended phase space, upon which the gauge currents are represented by Hamiltonian vector fields on the space of fields. Each of these approaches is a possible starting point for a quantum theory, some better than others presumably. 
Nonetheless, it is important to prepare the classical ground as suitably as possible for quantization.
The key point is to insure that all aspects of the geometry are included in the gravitational phase space, not only the spin-$2$ degrees of freedom but also the background geometry degrees of freedom that enters as we will see in the form of spin-$0$ and spin-$1$ canonical pairs. The spin-$0$ pair which plays a prominent role in our analysis consists of the area form conjugate to the surface tension, a boost connection encoding a combination of inaffinity and expansion.

In this paper, we will focus on the classical gravity problem on generic null hypersurfaces. There are certainly important examples of null hypersurfaces that have been extensively studied in the literature. This includes the null conformal boundary in asymptotically flat spacetimes; in this context the reduced phase space was long ago understood beginning with Ashtekar and Streubel  \cite{Ashtekar1981}. Another important example of a null hypersurface that has been much studied is a Killing horizon \cite{Wall:2011hj}. By a generic null hypersurface, we mean one that might be located anywhere in a spacetime without simplifying assumptions, and is not necessarily one of the two mentioned examples. In fact, one may consider 'segments' of hypersurfaces that are taken to end on specific corners, or such corners may be taken to be at asymptotic distance or at a (classical) singularity. Such a situation makes contact with the familiar discussions of the classical focusing of null congruences and the classical singularity theorems.

Gravity on null hypersurfaces, and in general null physics, possesses remarkable properties leading to dramatic simplifications (compared to space-like or time-like hypersurfaces) that we will demonstrate in this paper.
A lot of efforts in quantum gravity have been unsuccessfully focused on the  Wheeler-DeWitt equation which represents canonical quantization of constraints associated with spacelike surfaces \cite{Arnowitt:1962hi, dewitt1967quantum}. One of the challenges in this case is that the constraints are elliptic equations, whose solutions are non-local and no non-perturbative treatment is known. On the other hand,
constraints projected on null surfaces present themselves as equations of evolution along null time and are therefore  much simpler to solve non-perturbatively. The canonical structure localizes along one-dimensional null rays and they, therefore, provide a  more direct and simpler insight into the quantum nature of geometry.

A conceptual complication is that the geometry on a null surface is non-Riemannian. In particular, there is the degeneracy of the metric \cite{Henneaux1979a}, and the corresponding intricacies in projecting to the transverse and tangent subspaces \cite{Mars:1993mj, Gourgoulhon:2005ng, Gourgoulhon:2007ue}. In recent years, a superior understanding of null geometry has been achieved, with the realization that Carrollian physics plays a central role\cite{LevyLeblond1965, Gupta1966}. The null geometry is correspondingly understood in terms of data on a line bundle over a Euclidean space (which from a spacetime perspective is a corner) called a Carroll structure.  

In fact, it is now well-established how to encapsulate the symmetries and physical laws on null hypersurfaces \cite{Duval:2014uva, Duval:2014uoa, Duval:2014lpa, Donnay:2015abr, Donnay:2016ejv, Hopfmuller:2016scf, Penna:2018gfx,  Ciambelli:2018wre, Chandrasekaran:2018aop, Ciambelli:2019lap, Donnay:2019jiz, Ciambelli:2018ojf, Adami:2020amw, Chandrasekaran:2020wwn, Adami:2021kvx, Freidel:2022bai, Petkou:2022bmz}. One approach is to consider the induction of Carrollian geometry, including a metric, bundle structures and a connection, along with a symplectic structure, from a spacetime theory. Another equivalent approach is to formulate a Carrollian stress tensor directly on the null hypersurface. While this might be viewed as an effective 'fluid' perspective, we will in fact demonstrate below that these two approaches are completely equivalent. From the first point of view, there are important constraints that describe the properties of the null congruence (and from the Carroll perspective, the geometry of the Carroll structure) while from the latter point of view, the constraints correspond to the (broken) Ward identities of the Carrollian stress tensor. From each perspective, there are modes that we will refer to as spin-0, spin-1, and spin-2 fields, this nomenclature making reference to representation theory relevant to the codimension-2 base of the Carroll structure.  As just mentioned, in the latter perspective the Einstein equations correspond to Carrollian conservation equations. In particular, they correspond to the Raychaudhuri \cite{PhysRev.98.1123, Sachs:1961zz, Landau:1975pou} (see also \cite{Kar:2006ms})\footnote{The Raychaudhuri equation on null hypersurfaces was first derived in \cite{Sachs:1961zz}.} and Damour \cite{damour1978black, Damour1979} constraint equations, which involve all of the spin-0, spin-1, and spin-2 gravitational data as well as matter fields. The canonical charges can be constructed, and they reduce to codimension-$2$ charges on-shell of the constraints. 

The relationship between the intrinsic Carrollian presymplectic data and the induced gravitational presymplectic structure, off-shell of the constraints, has been studied in \cite{Chandrasekaran:2021hxc, Freidel:2022vjq}; see also \cite{Jafari:2019bpw} for an earlier proposal for the stress tensor on null boundaries. We here prove, using the framework of \cite{Freidel:2020xyx, Freidel:2020svx, Freidel:2020ayo}, that the bulk action, improved by a suitably chosen boundary Lagrangian, induces exactly the Carrollian canonical presymplectic potential, plus corner terms. Indeed, the final action is related to the one found in \cite{Lehner:2016vdi} by corner Lagrangians.

Building the gravitational presymplectic data on a generic null hypersurface is not an easy task, and the problem has attracted tremendous attention in recent years \cite{Parattu:2015gga, Parattu:2016trq, Lehner:2016vdi, Wieland:2017cmf, Wieland:2017zkf, Hopfmuller:2018fni, Oliveri:2019gvm, Adami:2020ugu, Adami:2021nnf, Chandrasekaran:2021hxc, Freidel:2022vjq, Sheikh-Jabbari:2022mqi, Adami:2023fbm}. The study of the characteristic initial value problem on null hypersurfaces was initiated much earlier by Sachs \cite{Sachs:1962zzb} and pursued in \cite{Gambini:1977yd, Penrose:1980yx,Torre:1985rw, Goldberg:1992st, Goldberg:1995gb, dInverno:2006wzl, Wieland:2019hkz, Wieland:2021vef, Mars:2022gsa, Mars:2023hty}  using also the constrained Hamiltonian formalism. The ensemble of works of Reisenberger stands out in this avenue of research \cite{Reisenberger:2007pq, Reisenberger:2007ku, Reisenberger:2012zq, Fuchs:2017jyk, Reisenberger:2018xkn}.

The main body of this manuscript is devoted to a full analysis of the presymplectic structure and Poisson bracket on a generic null surface in four spacetime dimensions. As we mentioned above, this involves an intricate interplay between spin-0,1,2 data. In this paper, we put much of our attention on the spin-0 and spin-2 contributions, as our intent is to focus on the Raychaudhuri constraint. The spin $0$ canonical data include the surface tension, which is often treated as a background structure. Here we see that it has to be understood dynamically as responding, through the constraint, to the presence of spin $2$ and matter degrees of freedom. 
It is profitable to formulate the spin-2 data (corresponding to the degenerate metric and shear) in terms of the Beltrami parameterization \cite{Beltrami}, the use of which in this context was first implemented by Reisenberger; see also \cite{Baulieu:2021, Baulieu:2023wqb} for another recent use of the Beltrami parameterization in gravity. We can then invert the off-shell presymplectic structure and read the resulting kinematic Poisson brackets. 

More generally though, there is a non-trivial mixing between the spin-0 and spin-2 modes which means that we cannot simply decouple the spin $0$ modes as is assumed in a perturbative treatment. The Poisson brackets are local on the base,  and their non-locality is only along  the null fibre direction. We find that the constraint algebra 
closes only when both the spin-2 and spin-0 contributions are included. 
This means that one cannot treat the spin-2 gravitational radiation independently of the geometry of a generic null hypersurface. Our formulation is however non-perturbative in Newton's constant, and thus can be regarded as including gravitational backreaction.

The constraint algebra that we referred to above is of course off-shell as the constraints themselves have not been imposed. The Poisson brackets may thus be regarded as kinematic. In a classical theory we would then look for a way to pass to the physical subspace upon which the constraints are imposed. We will show that 
the Raychaudhuri constraint generating the null time reparametrizations can be understood as the  vanishing of a total CFT-like stress tensor, which is the sum of spin-$0$, spin-$2$, and matter contributions. 
 In a quantum theory, the details are of course different, the constraint becoming an operator. Given the nature of the Raychaudhuri  constraint, we expect anomalies to play a key role at the quantum level. In the conclusion, we will comment upon these options but leave details to future work. 

There is a particular linear combination of a diffeomorphism along $\ell$ and internal boost, which we denote $M'_{f}$ with $f$ a function on the null geometry, that has the effect of preserving the reduced space of the spin-0 and spin-2 modes. 
Interestingly, we find that in the classical theory there is a canonical transformation the result of which has the interpretation of measuring null time via a field variable that emerges from the spin-0 sector. There is a choice of such a clock, called {\it dressing time}, that completely relegates to the corner the spin-0 symplectic pair on-shell. In the special case of Killing horizons this clock coincides with affine time, but is generally different. Remarkably, the dressing time clock variable is canonically conjugate to the Raychaudhuri constraint. All other fields, including the spin-2 gravitational mode and arbitrary matter fields, are dressed in the sense of becoming invariant under diffeomorphisms in the dressing time frame. We thus arrive at an interpretation in which an {\it observer} (in this case, simply a clock) emerges directly from the gravitational degrees of freedom. In a quantum context, this is closely related to the crossed product construction of Ref. \cite{Witten:2021unn, Jensen:2023yxy, Klinger:2023tgi}. Since the clock is associated with the spin-0 modes, one might have naively thought that these were pure gauge and had no physical impact. However, we will see very explicitly that this is not the case -- they instead give rise to a clock which is dynamical throughout the bulk of the hypersurface but is conjugate to a constraint. Promoting the clock to be dynamical has another virtue, which is that we can solve the Raychaudhuri constraint non-perturbatively. 

The boost charge we construct is given by the corner area, which is automatically positive. In the dressing time only, it is further monotonic on-shell, if the classical null energy condition is imposed. When quantum matter is added to the system, this charge represents therefore a good candidate to describe the generalized entropy \cite{Wall:2012uf}, and extend the quantum focusing conjecture \cite{Engelhardt:2014gca, Bousso:2015mna, Bousso:2015wca} to arbitrary null hypersurfaces. 

The raison d'\^etre of this manuscript is thus to appreciate that null hypersurfaces have surprising properties, leading to simplifications in dealing with gravity: the gravitational data can be fully recast as intrinsic Carrollian data, the symplectic structure can be inverted, the constraint can be solved non-perturbatively and is conjugate to the dressing time, it can be further understood as a balance equation for CFT-like stress tensors associated to the spin-$0$, spin-$2$ and matter sectors, the boost charge is positive and monotonic. All these features are essential in preparing the ground for future explorations. The long-term idea is that null physics offers a non-perturbative window to study quantum gravity from a bottom-up approach.

The paper is organized as follows. We begin in Section \ref{S2} by reviewing the salient ingredients of the geometry \ref{S21} and dynamics \ref{S22} of null hypersurfaces. Then, Section \ref{S3} is devoted to the canonical presymplectic phase space. From the general analysis \ref{S31} we  find a set of assumptions \ref{S32} isolating the spin-0 and spin-2 contributions. We then show how to relate this intrinsic analysis to the bulk gravitational Lagrangian \ref{S33}. In Section \ref{S4} we construct the Poisson brackets of the dynamical fields. The key ingredients are the complex structure \ref{S41} and the Beltrami parameterization \ref{S42}. They lead us to the notion of propagators \ref{S43}, which are instrumental in our analysis. We use them in \ref{S44} to compute the constraint algebra, and in particular the highly non-trivial mixing between the spin-0 and spin-2 Hamiltonians. Moving from an arbitrary coordinate null time to the dressing time, we show in Section \ref{S5} that the spin-$0$ data are pushed to the corner, and are replaced by the full constraint operator in the bulk. We conclude showing in \ref{sec53} that the boost charge is positive and monotonic in the dressing time frame. 
We then offer some outlook in Section \ref{Conclusions}. Technical details supporting the results of Section \ref{S4} appear in  appendixes \ref{A1}, \ref{A2}, and \ref{A3}. Similarly, appendix \ref{appendixclocksymp} contains useful materials for Section \ref{S5}.

\section{Null Hypersurfaces}\label{S2}

We review here the salient ingredients of the geometry and dynamics of null hypersurfaces. The main references are \cite{Ciambelli:2019lap, Chandrasekaran:2021hxc, Freidel:2022vjq}, see also \cite{Freidel:2023bnj}.

\subsection{Geometry}\label{S21}

An $(n+1)$-dimensional null hypersurface $\cN$ is intrinsically described as a manifold endowed with a \textit{Carrollian structure} $(q_{ab},\ell^b)$. It consists of a nowhere-vanishing vector field $\ell=\ell^a\pa_a$ and a rank-$n$ metric $q_{ab}$ such that $\ell^a q_{ab}=0$. Here, for simplicity, we will take $n=2$ (corresponding to a spacetime dimension $d=n+2=4$).
Any such Carrollian structure is unique up to a local internal boost sending $\ell$ to $\mathrm{e}^\lambda\ell$. Mathematically this means that $\mathcal{N}$ is a fibred manifold with fibration $\pi: \mathcal{N} \to S $, such that the null direction $\ell$ is in the kernel of $\rd \pi$ and $S$ is a Riemann surface that we will usually think of as a $2$-sphere.

A generic null hypersurface in a manifold has creases and caustics, see e.g. \cite{Siino:2004xe,Gadioux:2023pmw}. The Carrollian structure is applicable only to portions of null hypersurfaces in which creases and caustics are not present. In this regard, we are performing a quasi-local analysis on $\cN$, and thus we assume that $\cN$ is an open set with topology $S\times I$. 

We denote $\cC$ an arbitrary transverse cut, i.e., a closed codimension-$1$ surface on $\cN$ with non-degenerate induced metric. The degenerate metric $q_{ab}$ on $\cN$ determines a spatial area form $\ve_{\cC}$  such that $\iota_{\ell} \ve_{\cC}=0$, where $\iota_{\ell}$ is the interior product. Moreover, the choice of $\ell$ determines a volume form $\ve_{\mathcal{N}}$ on $\mathcal{N}$, such that $\iota_{\ell}\ve_{\mathcal{N}}= \ve_{{\cC}} $.
The expansion $\theta$, defined below, relates the two forms through $ \rd\ve_{{\cC}}=\theta \ve_{\mathcal{N}}$. 
In the following, we will assume that the manifold $\cN$ is such that its completion $\overline{\cN}$ has boundaries (corners)  $\partial \overline{\cN}=\cC_\pa$ with $\cC_\pa=\cC_{\mathsf{past}}\bigcup \cC_{\mathsf{future}}$. 

To construct a connection on $\mathcal{N}$ and  describe  dynamics on a null hypersurface, it is necessary to introduce an Ehresmann connection: an Ehresmann connection is a 1-form $k=k_a \rd x^a$ dual to the null vector $\ell$, $\iota_\ell k=1$. In the context of Carrollian physics, this was described in \cite{Bekaert:2015xua, Ciambelli:2019lap}. 
The Ehresmann connection defines a notion of horizontality, where $Y$ is a horizontal vector field on $\mathcal{N}$ if $\iota_Y k=0$. A general vector $\xi \in T\mathcal{N}$ can be decomposed as $\xi=f\ell + Y$, with $Y$ horizontal. With the Ehresmann connection one can decompose the volume form on $\mathcal{N}$ as $\ve_{\mathcal{N}}=k\wedge \ve_{\cC}$, and define the horizontal projector $q_a{}^b= \delta_a{}^b - k_a\ell^b$ satisfying $\ell^a q_a{}^b=0=q_a{}^b k_b$ and $q_a{}^bq_{bc}=q_{ac}$. Such a construction has been introduced in \cite{Mars:1993mj}; see also \cite{Gourgoulhon:2005ng}. The Carrollian literature is vast. Its link with BMS symmetries of null hypersurfaces, and their gravitational data  has been appreciated in \cite{Duval:2014uva, Duval:2014uoa, Duval:2014lpa, Donnay:2015abr, Donnay:2016ejv, Hopfmuller:2016scf, Penna:2018gfx,  Ciambelli:2018wre, Chandrasekaran:2018aop, Ciambelli:2019lap, Donnay:2019jiz, Ciambelli:2018ojf,  Adami:2020amw, Chandrasekaran:2020wwn, Adami:2021kvx, Freidel:2022bai}.

The set of data $(q_{ab}, \ell^b, k_a)$ defines a \emph{ruled Carrollian structure} on $\mathcal{N}$. The Carrollian acceleration $\varphi$ and vorticity $w$, which are a transverse 1-form and 2-form, are defined via
$\rd k=\varphi\wedge k+w$.
Setting the vorticity to zero corresponds to a choice of foliation for $\mathcal{N}$, that is, a global section of the bundle.
With the ruled Carrollian structure we can uniquely raise indices of horizontal tensors. As an illustrative example, given the expansion tensor, defined as the Lie derivative along $\ell $ of the degenerate metric,
\be 
\theta_{ab}\coloneqq  \frac12 {\cal L}_\ell q_{ab},
\ee 
the tensor $\theta_a{}^b$ is defined as the unique tensor satisfying $\theta_a{}^c q_{cb}=\theta_{ab}$ and $\theta_a{}^b k_b=0$. We can then decompose the expansion tensor in terms of the expansion $\theta$ and shear $\sigma_a{}^{b}$ of the null surface:
\be 
\theta_{a}{}^b= \frac{1}{2} \theta q_{a}{}^b + \sigma_{a}{}^b, \quad \text{with} \quad \sigma_a{}^a=0.
\ee 

A ruled Carrollian structure also allows us to introduce the notion of  \emph{Carrollian connections} $D_a$
discussed in the literature in \cite{Ciambelli:2018xat, Chandrasekaran:2021hxc,  Freidel:2022vjq}.\footnote{As discussed in \cite{Freidel:2022vjq}, the  Carrollian connection can be chosen to be induced by the bulk Levi-Civita connection $\nabla_a$ through $D_a T_b{}^c := \Pi_a{}^{a'}\Pi_b{}^{b'} \nabla_{a'}T_{b'}{}^{c'}\Pi_{c'}{}^c$ following the  framework of \cite{Mars:1993mj}. Here, $\Pi_a{}^b=q_a{}^b+k_a\ell^b$ is the bulk Rigging projector, which becomes $\delta_a^b$ on the null hypersurface.} We choose $D_a$ to be a torsionless connection preserving the identity, that is, $D_a (q_b{}^c+k_b \ell^c)=0$. This Carrollian connection defines a $1$-form $\omega=\omega_a\rd x^a$ on $\mathcal{N}$ given by
\be \label{boostf}
D_a \ve_{\mathcal{N}} = -\omega_a \ve_{\mathcal{N}}.
\ee
This $1$-form can be decomposed into a transverse and horizontal part as $ \omega_a =\kappa k_a + \pi_a$, where $\pi_a$ is the H\'ajiček connection and $\kappa$ the inaffinity of $\ell$.\footnote{The symbol $\kappa$ is often used to indicate the surface gravity. However, inaffinity and surface gravity are in general different,  coinciding only for Killing horizons. For more details, see for instance \cite{Jacobson:1993pf}, where the inaffinity is denoted $\kappa_2$ while the surface gravity $\kappa_1$.} The $1$-form $\omega$ is a boost connection since, under the boost $\ell\to e^\lambda \ell$ and $k\to e^{-\lambda} k$, it transforms as $\omega \to \omega + \rd \lambda$. 
The boost-covariantized derivative of $\ell$ is horizontal and given by the expansion tensor
\be 
(D_a-\omega_a)\ell^b =\theta_a{}^b.
\ee 

The Carrollian connection chosen only preserves the metric when derivatives and tensor indices are taken to be horizontal. In general, when the expansion is non-vanishing the induced metric is not preserved by the connection, and one has 
\be 
D_a q_{bc}=- k_b \theta_{ac}- k_c \theta_{ab}.
\ee
To recapitulate, this Carrollian connection has been found requiring absence of torsion, and metricity in the horizontal directions. In particular, these properties guarantee that we can write the expansion tensor \`a la Brown-York, that is,
\beq
\theta_{ab}=\frac12 \cL_\ell q_{ab}= D_{(a}\ell^c q_{b)c}.
\eeq

\subsection{Dynamics}\label{S22}

We now turn our attention to the dynamics induced from Einstein's equations.
It is well-known that the projection of Einstein's equations along a null hypersurface leads to the Raychaudhuri \cite{PhysRev.98.1123} and Damour  \cite{damour1978black} equations, see also \cite{price1986membrane} and \cite{Donnay:2019jiz, Freidel:2022vjq}. These are given by ($R_{\ell\ell}=\ell^a R_{ab} \ell^b$)
\bea \label{Ray}
({\cal L}_\ell+\theta)[\theta] &=& \mu \theta -\sigma_{a}{}^b\sigma_b{}^a 
- R_{\ell \ell},\\ \label{Damour}
q_a{}^b \left({\cal L}_\ell+\theta\right)[\pi_b]+\theta\varphi_a &=& (\overline{D}_b+\varphi_b)(\mu q_a{}^b-\sigma_{a}{}^b)
+q_a{}^bR_{b\ell},
\eea
where we have introduced the {\it surface tension}\footnote{These equations can be extended to any spacetime dimension $d\geq3$. For example, one has $\mu= \kappa + \frac{d-3}{d-2}\theta$.} of $\mathcal{N}$
\be 
\mu :=  \kappa +\frac{\theta}{2}
\ee  
and $\overline{D}_a=q_a{}^b D_b$ denotes the horizontal derivative. We denote $\cL_\xi= \iota_\xi \rd + \rd \iota_\xi$ the spacetime Lie derivative along $\xi$. One can check that these equations are covariant under the boost symmetry $\ell \to e^{\lambda} \ell$ due to the fact that $\mu$ transforms as a boost connection $\mu \to e^{\lambda}(\mu +\ell[\lambda])$. We also note that the Carrollian vorticity $w$ does not contribute, while the Carrollian acceleration $\varphi$ contributes only to the Damour equation. The surface tension $\mu$ is an important datum in our construction, and will play a central role in the rest of the manuscript.

Equations \eqref{Ray} and \eqref{Damour} can alternatively be understood as conservation equations of a Carrollian fluid.   
One defines the Carrollian fluid energy-momentum tensor in terms of the rigged projector and the Carrollian connection  as 
\be 
8\pi G\,T_a{}^b \coloneqq D_a\ell^b - \delta_a{}^b D_c\ell^c.
\ee 
The tensor $D_a\ell^b =\omega_a \ell^b +\theta_a{}^b$ is the Weingarten map \cite{Gourgoulhon:2005ng}. The Carrollian fluid energy-momentum tensor can be decomposed in terms of the quantities defined above as $T_a{}^b =\tau_a \ell^b +\tau_a{}^b$, where 
\beqn
8\pi G\,\tau_a&=&  \pi_a-\theta k_a,\\
8\pi G\, \tau_a{}^{b}&=&\sigma_a{}^{b}-\mu q_a{}^{b}.
\eeqn

As shown in \cite{Chandrasekaran:2021hxc} (see also \cite{Freidel:2022bai,Freidel:2022vjq}), Einstein's equations pulled back on $\mathcal{N}$ can then simply be written as fluid conservation equations
\be \label{DT}
D_b T_a{}^b= T^{\mathsf{mat}}_{a\ell},
\ee 
where $T_{ab}^{\mathrm{mat}}$ is the bulk matter energy-momentum tensor. 
This generalizes to null surfaces the  Brown--York result \cite{brown1993quasilocal} and is in agreement with the membrane paradigm \cite{price1986membrane}. The Carrollian energy-momentum tensor can be seen as  the $c\to 0$ limit of a relativistic energy-momentum tensor. In this fluid analogy, $\theta$ is the fluid energy, $\pi_a$ the heat flux, $-\mu$ the pressure, $\sigma_{ab}$ the viscous stress tensor,  and
$T_{a\ell}^{\mathsf{mat}}$ the external energy flux \cite{Ciambelli:2018xat, Rezolla, Petkou:2022bmz}.

The Raychaudhuri equation \eqref{Ray} can be derived from \eqref{DT} by projecting the latter on $\ell$
\bea
8\pi G\,\ell^bD_aT_b{}^a
=-(\cL_{\ell}+\theta)[\theta]+\mu\theta-\sigma_a{}^b\sigma_b{}^a.
\eea
Similarly, the Damour equation is obtained by projecting \eqref{DT} on $q_c{}^b$, see \cite{Freidel:2022vjq}.

\section{Presymplectic Phase Space}\label{S3}

As shown in \cite{Hopfmuller:2018fni}, the Raychaudhuri and Damour equations (\ref{Ray},\ref{Damour}) can be understood as (non)-conservation equations of corner symmetry charges associated with diffeomorphisms along $\ell$ and time-dependent diffeomorphisms of the base $S$. We are going to review that the corresponding charge aspects are $-\theta \ve_{\cC}$ and $\pi_a\ve_{\cC}$, respectively. To see this, we first introduce the canonical presymplectic potential\footnote{In this section, we deal with the off-shell presymplectic phase space.} intrinsic to $\cN$ 
\beq\label{thcan}
\Theta^{\mathsf{can}}=\int_{\cN}\theta^{\mathsf{can}}=\int_{\cN} \ve_{\cN}\Big(\frac{1}{2}
\tau^{ab}\delta q_{ab}
-\tau_a\delta \ell^a\Big).
\eeq
This expression first appeared in \cite{Chandrasekaran:2021hxc}, and shows that $-\tau_a$ and $\tfrac12 \tau^{ab}$ are the momenta conjugate to the Carrollian vector $\ell=\ell^a\pa_a$  and  the degenerate Carrollian metric $q_{ab}$, respectively. The tensor $\tau^{ab}$ is defined as the unique symmetric tensor such that $q_{ac}\tau^{cb}=\tau_a{}^b$ and $\tau^{ab} k_b=0$. Note that while $q_{ab}$ is not invertible, by $q^{ab}$ we mean the unique symmetric tensor for which $q^{ab}k_b=0$. This satisfies $q_a{}^b=q_{ac}q^{cb}$. 

\subsection{Diffeomorphisms and Canonical Generators }\label{S31}

To evaluate the canonical charges associated with diffeomorphisms one uses that the integrand of the presymplectic potential  is manifestly invariant under diffeomorphisms. We introduce field space calculus $\fL_{\hat{\xi}}=I_{\hat{\xi}}\delta+\delta I_{\hat\xi}$, with $\hat\xi$ the Hamiltonian vector field associated with the diffeomorphism $\xi\in T\cN$; see \cite{Ciambelli:2022vot, Ciambelli:2023bmn} for details.  To proceed, we study how the fields transform. Since at this stage no assumptions have been made on field space, we have that the fields transform covariantly via the Lie derivative
 \beq
\fL_{\hat{\xi}} q_{ab} &=&\xi^c D_c q_{ab}+q_{ac}D_b \xi^c+q_{cb}D_a\xi^c\\
\fL_{\hat{\xi}}\ell^a &=& \xi^b D_b\ell^a-\ell^b D_b \xi^a\\
\fL_{\hat{\xi}} \tau^{ab} &=&\xi^c D_c \tau^{ab}-\tau^{ac}D_b \xi^c-\tau^{cb}D_a\xi^c\\
\fL_{\hat{\xi}}\tau_a &=& \xi^b D_b\tau_a+\tau_b D_a \xi^b.
 \eeq
We then find that  
$\fL_{\hat{\xi}}\Theta^{\mathsf{can}}= \cF^{\mathsf{can}}_\xi$, where the canonical flux $\cF^{\mathsf{can}}_\xi$  is a corner term given by 
\be \label{fff}
\cF^{\mathsf{can}}_\xi= \int_{\cC} \iota_{\xi}\ve_{\cN}\Big(\frac{1}{2}\tau^{ab}\delta q_{ab}-\tau_a\delta \ell^a\Big).
\ee
Here, and in the following, we denote $\cC$ the boundary of $\cN$, previously called $\cC_\pa$. The integrand of the flux is a spacetime 2-form and field space one-form simply given by the contraction $\iota_\xi \theta^{\mathsf{can}}$. The canonical flux vanishes when $\xi|_{\cC}\in T\cC$, or when the phase space is properly extended \cite{Ciambelli:2021nmv,Freidel:2021dxw, Klinger:2023qna}.

The charge is then extracted from
\be 
I_{\hat{\xi}}\Omega^{\mathsf{can}} = -\delta Q_\xi+\cF_\xi^{\mathsf{can}} \qquad Q_\xi=I_{\hat\xi}\Theta^{\mathsf{can}},
\ee
which shows that $I_{\hat{\xi}} \Theta^{\mathsf{can}}$ is the canonical charge for diffeomorphism symmetry.
This result holds assuming  $\delta \xi=0$.

We therefore find that the canonical charge is the sum of the constraint (the equations of motion) plus the corner charge
\bea\label{ixcan}
I_{\hat{\xi}} \Theta^{\mathsf{can}}&=&
\int_{\cN}\ve_{\cN} \Big(T_a{}^b D_b\xi^{a}-\tau_a \xi^b D_b\ell^a \Big)\nonumber\\
&=&
\int_{\cN}\ve_{\cN} \Big(T_a{}^b (D_b-\omega_b)\xi^{a}\Big)\cr&=& 
-\int_{\cN} \ve_{\cN}\;\xi^{a} D_bT_a{}^b 
+  \int_{\cC} \ve_{b}\;\xi^{a}T_a{}^b,  
\eea
where we used $ \tau_bD_a \ell^b = T_a{}^b \omega_b$ in the second equality, eq. \eqref{boostf} in the third, and we denoted $\ve_b =\iota_{\partial_b} \ve_{\mathcal{N}}$.
The first term in the last equality of \eqref{ixcan} vanishes on-shell,\footnote{The derivation is given for pure gravity. If matter is present, there is another contribution $\int_{\cN}\ve_{\cN}
\xi^a T^{\mathsf{mat}}_{a\ell}= I_{\hat{\xi}}\Theta^{\mathsf{mat}}$ to the first term, and so the first term generally vanishes on-shell.} showing that the charges associated with a vector field $\xi$ tangent to $\mathcal{N}$ are corner charges given by 
\be \label{charg}
Q_\xi = \int_{\cN}\ve_{\cN}\, T_a{}^b (D_b-\omega_b)\xi^{a}  \,\hat{=}\,
 \int_{\cC} \ve_b\,\xi^aT_a{}^b ,
\ee 
where $\hat{=}$ denotes the on-shell evaluation. 

From now on, we choose the transverse cut $\cC$ to be such that  $\ve_b$ pulls back to $k_b\ve_{\cC}$. The diffeomorphism $\xi$ is given by an arbitrary\footnote{The time-independent and linear-in-time modes of $f$ are the supertranslation and superboost parameters, respectively.} diffeomorphism $f$ along $\ell$ and a transverse diffeomorphism, $Y^a$, such that\footnote{We continue to require the  field independence of the vector field,
i.e., $\delta\xi=0$, and thus one has to select $f$ and $Y^a$ such that this holds.}
\beq
\xi^a=f\ell^a+Y^bq_b{}^a .
\eeq

Let us compute the charge $Q_{f\ell}=M_f$ associated to  $\xi^a=f \ell^a$. We have
\beq\label{Mf}
M_f=\frac1{8\pi G} \int_{\cN}\ve_{\cN}\left( f\sigma_a{}^b\sigma_b{}^a-\theta (\ell+\mu)[f] \right)\hat{=}-\frac1{8\pi G} \int_{\cC} \ve_{\cC}\, f \theta.
\eeq
The last step could have been derived directly from
\eqref{charg}, since $\ell^a T_a{}^b \ve_b=-\theta \ve_{\cC}$. On the other hand, for a diffeomorphism along ${\cC}$, $\xi^a=q_b{}^a Y^b$, we have ($Q_{Y}=P_Y$)
\bea
P_Y &=&\frac1{8\pi G}\int_{\cN}\ve_{\cN}\Big(\sigma_c{}^b(D_b-\pi_b)Y^c+\pi_c(D_\ell-\frac{\theta}{2})Y^c-\mu \overline{D}_cY^c-\theta Y^c\varphi_c\Big)\\
&\hat{=}&\frac1{8\pi G} \int_{\cC}\ve_{\cC}\, Y^a\pi_a ,
\eea
where $D_\ell=\ell^aD_a$. As expected, the charges become corner charges on-shell.

If the  canonical flux vanishes at the corner $\cF^{\mathsf{can}}_\xi=0$, which is equivalent to the condition $f|_\cC=0$, one can simply derive the charge algebra from the Noether analysis. Indeed,  under this assumption $\delta Q_\xi=I_{\hat{\xi}} \Omega^{\mathsf{can}} $ and we have
\bea\label{conv}
\{Q_\xi,Q_\zeta\}=\Omega^{\mathsf{can}}(\hat{\xi},\hat{\zeta}) =
-I_{\hat{\xi}} I_{\hat{\zeta}} \Omega^{\mathsf{can}}  =I_{\hat{\xi}} \delta I_{\hat{\zeta}}\Theta^{\mathsf{can}} =
\fL_{\hat{\xi}} I_{\hat{\zeta}}\Theta^{\mathsf{can}}= 
I_{[\hat{\xi},\hat{\zeta}]}\Theta^{\mathsf{can}}=-Q_{[{\xi},{\zeta}]},
\eea
where the last equality is a consequence of $[\hat{\xi},\hat{\zeta}]=-\widehat{[{\xi},{\zeta}]}$.
This means that 
\bea \label{masscon}
\{M_f,M_g\}&=&-M_{(f\ell[g]-g\ell[f])}\\
\{P_Y,M_f\}&=& -M_{Y[f]}\\
\{P_Y, P_{Y'}\}&=& -P_{[Y,Y']_{\mathrm{Lie}}}.
\eea
We note that one could have kept the dissipative terms and instead computed the charge algebra using the extended phase space introduced in  \cite{Ciambelli:2021nmv,Freidel:2021dxw, Klinger:2023qna}.
\subsection{Boost Symmetry}

As already noticed, Carrollian fluids possess another symmetry, which is the internal local boost symmetry generated by the infinitesimal parameter $\lambda$. Given the canonical presymplectic potential, we can compute the associated charges. This symmetry acts on the fields as
\beq
\fL_{\hat\lambda} \ell=\lambda \ell, \quad \fL_{\hat\lambda} \epsilon_{\cN}=-\lambda \epsilon_{\cN} \quad \fL_{\hat\lambda} q_{ab}=0,\quad \fL_{\hat\lambda}\theta=\lambda \theta,\quad \fL_{\hat\lambda}\mu=\ell(\lambda)+\lambda\mu.
\eeq
From this, we derive
\beq
\fL_{\hat\lambda} \tau^{ab}=\lambda \tau^{ab}-\frac1{8\pi G}\ell(\lambda)q^{ab}, \quad \fL_{\hat\lambda} \tau_a=\frac1{8\pi G}\overline{D}_a\lambda.
\eeq

To compute the charge, assuming from now on $\delta \lambda=0$, we start by evaluating
\beq
\fL_{\hat\lambda}\Theta^{\mathsf{can}}=-\frac1{8\pi G}\int_{\cN}\ve_{\cN}\Big(\frac{1}{2}\ell(\lambda)q^{ab}\delta q_{ab}+\overline{D}_a\lambda\delta\ell^a\Big)= -\frac1{8\pi G}\delta\Big( \int_{\cN}\ve_{\cN}\,\ell(\lambda)\Big),
\eeq
where we used that $\delta \ve_{\cN} = (\frac12 q^{ab}\delta q_{ab} - k_a \delta \ell^a)\ve_{\cN}$.
We also have 
\beq
I_{\hat\lambda}\Theta^{\mathsf{can}}=\frac1{8\pi G}\int_{\cN}\ve_{\cN}\,\lambda\theta.
\eeq
Then, we find that the boost  charge is a corner charge even off-shell
\beq
I_{\hat\lambda}\Omega^{\mathsf{can}}=\fL_{\hat\lambda}\Theta^{\mathsf{can}}  -\delta I_{\hat\lambda}\Theta^{\mathsf{can}} =- \frac1{8\pi G}\delta\left(\int_{\cC} \ve_{{\cC}}\,\lambda \right).
\eeq
This means that the boost charge is given by the corner area element
\beq
G_\lambda =\frac1{8\pi G}\int_{\cC}\ve_{\cC}\, \lambda .
\eeq
The charge algebra derived earlier can therefore be extended by the boost generator to give
\bea \label{masscon2}
\{ P_{Y}, G_\lambda\}&=& -G_{Y[\lambda]} \\
\{M_f, G_\lambda\}&=& - G_{f\ell[\lambda]}.
\eea

\subsection{Simplifying Assumptions}\label{S32}

In this paper, we will make a set of simplifying assumptions that will bring us to a framework suitable to study the Raychaudhuri equation.  
We start by denoting $\Omega:= \sqrt{\mathrm{det}(q)}>0$ the spatial measure factor. We also choose adapted coordinates $x^a=(v,z,\bz)$, where $(z,\bz)$  are complex coordinates for  the base of the Carrollian bundle and such that 
\beq
\ell=e^{-\alpha}\pa_v,
\eeq
where $\alpha$  labels the time scale.
We will restrict our analysis to variations that do not change the Carrollian structure:
\beq\label{dl}
\delta \ell=-\delta \alpha\, \ell.
\eeq
In these coordinates, the measure and expansion on $\cN$ are given by 
\be 
\ve_{\cN} = e^\alpha \Omega \ve_{\cN}^{(0)}, \qquad \ve_{\cC}= \Omega \ve^{(0)}_{{\cC}}, \qquad 
\theta = e^{-\alpha} \pa_v\log \Omega,
\ee 
where we recall that $\ve_{{\cC}}=\iota_{\ell}\ve_{\mathcal{N}}$ and we introduced the bare measure $\ve_{\cN}^{(0)}=\rd v\wedge \ve^{(0)}_{{\cC}}$ with $\ve^{(0)}_{{\cC}}=\rd^2 z $.
The variation of the measure is then
\beq
\delta \ve_{\cN}=\delta\alpha\; \ve_{\cN}+\delta\log\Omega\; \ve_{\cN} \qquad \delta\ve_{\cC}=\delta\log\Omega\;\ve_{\cC}.
\eeq

While the spin-$1$ contribution drops with these assumptions, the spin-$0$ and spin-$2$ contributions to the presymplectic potential read
\beq
\Theta^{\mathsf{can}}=\frac{1}{8\pi G}\int_{\cN}\ve_{\cN}
\Big(\tfrac12\sigma^{ab}\delta q_{ab} -\theta \delta \alpha-\mu \delta \log\Omega\Big).
\eeq
The presymplectic 2-form $\Omega^{\mathsf{can}}$ can be written more simply if one rewrites everything with respect to the rescaled null generator $ \tilde{\ell}:=\pa_v= e^\alpha \ell$. The associated surface tension is $\tilde{\mu}= e^\alpha \mu +\pa_v\alpha$, while the associated shear is $\tilde\sigma^{ab}=e^{\alpha}\sigma^{ab}$. Introducing the measure $\tilde\ve_{\cN}=\rd v\wedge \ve_{\cC}=e^{-\alpha}\ve_{\cN}$, we have\footnote{Once rewritten in terms of the rescaled data, there is an extra exact variational term in $\Theta^{\mathsf{can}}$ given by $-\delta\left(\int_{\cN} \ve_{\cN}\, \alpha \,\pa_v\log\Omega\right)$, which will vanish with our assumption $\alpha=0$.}
\beqn
\Omega^{\mathsf{can}}=\frac{1}{8\pi G}\int_{\cN}\tilde\ve_{\cN} \Big(-\delta \tilde{\mu}\wedge \delta \log\Omega+\tfrac1{2\Omega}  \delta ( \Omega \tilde\sigma^{ab})\wedge\delta q_{ab}\Big)+ \frac{1}{8\pi G}\int_{\cC} \ve_{\cC}\left(\delta \alpha \wedge \delta \log \Omega\right) .
\eeqn
The variation $\delta\alpha$ appears only in the last term, which is a corner contribution. The corner contribution and physics is ultimately crucial in order to understand gravity and edge modes. However, we first need to solve the constraints on the bulk of the hypersurface. 

We thus want to impose $\delta \alpha=0$, and further $\alpha=0$. This allows us to identify the Carrollian null vector $\ell$ simply with $\pa_v$, and require the symmetries to preserve this notion of time. 
To systematically achieve this, we must require the symmetry transformations to be such that $\delta\alpha=0$. We note that the assumption \eqref{dl} is preserved only by a transverse diffeomorphism $Y$ that does not depend on the coordinate $v$, such that $[\ell,Y]=0$. This selects among all diffeomorphisms on $\cN$ those that are automorphisms of the Carrollian bundle. Such symmetries have been called Carrollian diffeomorphisms in \cite{Ciambelli:2018xat, Ciambelli:2018wre, Ciambelli:2019lap}.

With this restriction, we recall that the symmetries act on $\alpha$ as
\beq
\fL_{\hat f}\alpha=e^{-\alpha}\pa_v f=\pa_v f \qquad \fL_{\hat \lambda}\alpha=-\lambda \qquad \fL_{\hat Y}\alpha=0,
\eeq
where we used that $\alpha=0$. We have also introduced the pragmatic notation $\fL_{\hat f}=\fL_{\hat \xi}$ with $\xi=f\pa_v$ and $\fL_{\hat Y}=\fL_{\hat \xi}$ with $\xi=Y$. We remark that,
imposing $\alpha=0$, the diffeomorphism along $\ell$ is explicitly field-independent,
$\xi=f\ell=f\pa_v \label{fpa}$.

We therefore see that in order to preserve the condition $\delta\alpha=0$ we need to combine the action of the diffeomorphism along $\ell$ with  a specific boost generated by $\lambda_f=\pa_v f$:
\beq\label{combinedsupertrans}
\fL'_{\hat{f}}
: =\fL_{f} + \fL_{\lambda_f}.
\eeq
This transformation is such that $\fL'_{\hat{f}}\alpha=0$ and thus $\fL'_{\hat{f}}\ell =0$. 

The combined transformation acts on the field space as
\beq\label{trs}
\fL'_{\hat{f}}q_{ab}=\cL_f q_{ab},\qquad
\fL'_{\hat{f}} \sigma_a{}^b =   \pa_v( f\sigma_a{}^b)\qquad\qquad\\
\fL'_{\hat{f}}\mu=\pa_v((\pa_v+\mu)f)\label{mutr},\qquad
\fL'_{\hat{f}}\Omega=f\pa_v\Omega\label{omtr},\qquad
\fL'_{\hat{f}}\theta=\pa_v(f\theta)\label{lft}
. \label{sitr}
\eeq
To summarize, we have shown that there is a consistent way to set $\delta\alpha=0$ by combining diffeomorphisms along $\ell$ and boosts. These two symmetries cease to exist as independent transformations on this reduced phase space. They combine together and act via $\fL'$, while the transverse diffeomorphism is unaffected.
 To stress the reduction, we will also use the notation $\fL'_{\hat Y}$, which is simply equal to $\fL_{\hat Y}$. It is important to stress that under this transformation the surface tension $\mu$ transforms anomalously, i.e. as a connection, not a scalar. We remark that, given \eqref{mutr}, if the surface tension $\mu$ is zero, it would remain zero on the phase space if $f$ is at best linear in $v$. This selects among the Carrollian diffeomorphisms the Weyl BMS subgroup of symmetries studied in \cite{Freidel:2021fxf}.

Dropping the tilde notation, with these restrictions the canonical presymplectic $2$-form reduces to
\beqn\label{2fo}
\Omega^{\mathsf{can}}=\frac{1}{8\pi G}\int_{\cN}\ve_{\cN} \Big(-\delta \mu\wedge \delta \log\Omega+\tfrac1{2\Omega}  \delta ( \Omega \sigma^{ab})\wedge\delta q_{ab}\Big),
\eeqn
where all quantities are now computed using the null generator $\ell=\pa_v$. In particular
\beq\label{sigma}
\sigma^{ab}=\frac12q^{aa'}q^{bb'}\left(\pa_v q_{a'b'}- \theta q_{a'b'} \right) = \frac{\Omega}2 q^{aa'}q^{bb'}  \pa_v \left(\frac{q_{a'b'}}{\Omega}\right) .
\eeq
This is our starting point for the kinematic Poisson bracket, in the next section. 

We compute the canonical charges directly from the presymplectic $2$-form
\beq
I'_{\hat f}\Omega^{\mathsf{can}}&=&\frac{1}{8\pi G}\int_{\cN}\ve^{(0)}_{\cN} \Big(-\fL'_{\hat{f}} \mu \delta \Omega+\tfrac1{2}  \fL'_{\hat{f}} ( \Omega \sigma^{ab})\delta q_{ab}+ \delta\mu\fL'_{\hat{f}} \Omega-\tfrac1{2} \delta(\Omega \sigma^{ab})\fL'_{\hat{f}}q_{ab}\Big)\\
&=&\frac{1}{8\pi G}\int_{\cN}\ve^{(0)}_{\cN} \Big(\delta\mu f\pa_v\Omega-(\pa_v(\pa_v+\mu)(f)) \delta \Omega-\delta(\Omega f \sigma_{a}{}^{b}\sigma_{b}{}^a)+\tfrac1{2}  \pa_v(f\Omega \sigma^{ab}\delta q_{ab})\Big)\\
&=&\frac{1}{8\pi G}\int_{\cN}\ve^{(0)}_{\cN} \Big(\delta(-\pa^2_vf  \Omega+\mu f\pa_v\Omega-\Omega f\sigma_{a}{}^{b}\sigma_{b}{}^a)+\pa_v(\tfrac1{2}f\Omega \sigma^{ab}\delta q_{ab}-f\mu\delta\Omega)\Big),
\eeq
where in the second step we used \eqref{sigma}. We can rewrite this as
\beq
I'_{\hat f}\Omega^{\mathsf{can}}=-\delta M'_f+\cF_f,
\eeq
where the last term is the expected flux $\cF_{f}:=\int_{\cC}f \iota_{\pa_v} \theta^{\mathsf{can}}$, while the first term gives the new  charge
\beq\label{primech}
M'_f = \frac1{8\pi G}\int_{\cN}  \ve^{(0)}_{\cN} f C + \frac1{8\pi G}\int_{\cC}\ve^{(0)}_{\cC}\left(\Omega \pa_v f-f\pa_v \Omega\right),
\eeq
where $C:= \pa_v^2\Omega-\mu\pa_v\Omega + \Omega \sigma_{a}{}^{b}\sigma_{b}{}^a$ is the Raychaudhuri constraint. 
As a consistency check, we note that the new charge is simply the sum of the old charge associated to a diffeomorphism along $\ell$ and the internal boost charge: $ M'_f=M_f + G_{\pa_v f}$.
The transverse diffeomorphism charges are unchanged, $P'_Y=P_Y$  when $Y$ is Carrollian.

In the absence of fluxes, we compute that the charge algebra
\beq
\{M'_f,M'_g\}&=&-M'_{(f\ell[g]-g\ell[f])}
\\
\{P'_Y,M'_f\}&=& -M'_{Y[f]}\\
\{P'_Y, P'_{Y'}\}&=& -P'_{[Y,Y']_{\mathrm{Lie}}}
\eeq
has the same structure as before, as expected.

\subsection{Bulk Lagrangian}\label{S33}

So far, we have confined our attention to the intrinsic Carrollian structure on null hypersurfaces. We have shown that the Carrollian conservation laws -- the Carrollian equations of motion -- encode the Einstein constraints, once projected onto the null hypersurface. But to proceed, we also need to show that the presymplectic structure is the same, as we are interested in the kinematic Poisson brackets of the dynamical data. 

To do so, consider the $4$-dimensional Einstein-Hilbert action coupled to matter 
\beq
L_{\text{EH}}=\left(\frac{1}{16\pi G}R+L^{\mathrm{mat}}\right)\ve,
\eeq
where $\ve = \sqrt{|g|}\rd^4x$ is the volume form, and we put $c=1$.
Its local presymplectic potential  \be \theta_{\text{EH}}= \frac{1}{16\pi G} \nabla_{\mu} (g^{\nu\sigma} g^{\mu\rho}\delta g_{\sigma\rho} - g^{\nu\mu} g^{\rho\sigma}\delta g_{\rho\sigma}) \ve_{\nu}\ee
is found through the variation $\delta L_{\text{EH}} = -E+\rd \theta_{\text{EH}}$, where $E = \frac{1}{16\pi G}\left(8\pi G T^{\mathsf{mat}}_{\mu\nu}-G_{\mu\nu}\right) \delta g^{\mu\nu}\ve$ are the equations of motion and $T^{\mathsf{mat}}_{\mu\nu}=\frac{-2}{\sqrt{|g|}}\frac{\delta (\sqrt{|g|}L^{\mathsf{mat}})}{\delta g^{\mu\nu}}$.

Consider a null hypersurface $\cN$, and introduce coordinates such that the Carrollian vector on ${\cal N}$ 
and the $1$-form normal to ${\cal N}$ are given by\footnote{The quantity $\bar{\alpha}$ ensures the normalization condition $\iota_k n=1$.}
\beq
\ell=e^{-\alpha}\pa_v, \qquad n=e^{\bar\alpha}\rd r.
\eeq
In \cite{Freidel:2022vjq}, it has been shown that the Einstein-Hilbert presymplectic potential is related to the Carrollian canonical presymplectic potential \eqref{thcan} as\footnote{While we assume here that the spin-$1$ field ($V^A$ in \cite{Freidel:2022vjq}) vanishes, this equation holds in full generality.}
\beq\label{theh}
\theta_{\text{EH}}=\theta^{\mathsf{can}}+\delta \ell_{\cN}-\rd \theta_{\cC},
\eeq
where we defined
\beq
\ell_{\cN}=\frac1{8\pi G}
\ve_{\cN}(\kappa+\theta),\qquad \theta_{\cC}=\frac1{16\pi G}
\ve_{\cC}\delta(\bar{\alpha}-\alpha).
\eeq
We recall that $\kappa$ is the inaffinity, which is related to $\alpha$ and $\bar\alpha$ via $\kappa=\ell(\bar\alpha)+\cdots$, where the dots represent terms linear in the fields that do not contribute to the presymplectic potential associated to $\ell_{\cN}$, and thus are not relevant in the following.

Our goal is to show that the bulk Einstein-Hilbert action can be improved by a boundary Lagrangian, such that the full bulk potential is equivalent to the  Carrollian potential plus a corner contribution.

Let us review the procedure outlined in \cite{Freidel:2020xyx, Freidel:2020svx, Freidel:2020ayo}. Essentially, two bulk Lagrangians differing by a boundary Lagrangian
\beq
L_1=L_2-\rd \ell_{\mathsf{bdy}},
\eeq
have symplectic potentials related via
\beq
\theta_1=\theta_2-\delta\ell_{\mathsf{bdy}}+\rd \theta_{\mathsf{bdy}}.
\eeq
where $\theta_{\mathsf{bdy}}$ is the symplectic potential of $\ell_{\mathsf{bdy}}$.

Given the result \eqref{theh}, it is natural to define an improved bulk Langrangian given by
\beq
L^{\mathsf{tot}}=L_{\text{EH}}-\rd \ell_{\cN}.
\eeq
To compute $\theta^{\mathsf{tot}}$, we thus need to evaluate the corner potential $\theta_{\cN}$. Using $\delta\ell_{\cN}=-E+\rd \theta_{\cN}$, and since $\kappa +\theta =\ell(\bar\alpha + \ln \Omega)+\cdots$ where the dots are terms with no time derivatives, we explicitly find
\beq
\theta_{\cN}=\frac1{8\pi G}\left(\delta\bar\alpha+\Omega^{-1}\delta\Omega\right) \ve_{\cC}.
\eeq

Therefore the total presymplectic potential becomes
\beq
\theta^{\mathsf{tot}}&=&\theta_{\text{EH}}-\delta \ell_{\cN}+\rd \theta_{\cN}\\
&=&\theta^{\mathsf{can}}+\rd (\theta_{\cN}-\theta_{\cC})\\
&=&\theta^{\mathsf{can}}+\frac1{16\pi G}\rd \left(\ve_{\cC}(\delta\alpha+\delta\bar\alpha+2\Omega^{-1}\delta\Omega)\right)\\
&=&\theta^{\mathsf{can}}+\rd\delta\ell_{\mathsf{cor}}+\frac1{16\pi G}\rd (\ve_{\cC}(\delta\alpha+\delta\bar\alpha)),
\eeq
where we used \eqref{theh} in the second line and introduced the corner Lagrangian $\ell_{\mathsf{cor}}=\frac1{8\pi G}\ve_{\cC}$ in the last step.

Adding this corner Lagrangian to the total action, one exactly retrieves the action found in \cite{Lehner:2016vdi}, when $\alpha=0$, studied further in \cite{Adami:2021nnf}. The boundary Lagrangian is the null analogue of the Gibbons-Hawking term, \cite{Parattu:2015gga, Parattu:2016trq}, while the corner term is the null analog of the Hayward term \cite{Hayward:1993my}. The symplectic $2$-form $\Omega^{\mathsf{tot}}$ is
\beq
\Omega^{\mathsf{tot}}=\delta\Theta^{\mathsf{tot}}=\Omega^{\mathsf{can}}+\frac1{16\pi G}\int_{\cC}\delta\ve_{\cC}\wedge \delta(\alpha+\bar\alpha).
\eeq
This exactly matches \eqref{2fo}, under our assumptions on $\alpha$ and further requiring $\delta\bar\alpha\vert_{\cC}=0$. This shows that the bulk Einstein-Hilbert action can be improved by a boundary Lagrangian, such that the final gravitational presymplectic structure on a null hypersurface is exactly the intrinsic Carrollian presymplectic structure.

\section{Kinematic Poisson Bracket}\label{S4}
In this section we compute the kinematic\footnote{This means the bracket before imposing the constraints.} Poisson bracket of the Carrollian presymplectic phase space variables. Doing so requires performing integration by part, so in this section we assume that all variations vanishes on $\pa \cN$.

\subsection{Unimodular Decomposition and Complex Structure}\label{S41}

Our starting point is the canonical presymplectic $2$-form \eqref{2fo}, that we rearranged into
\beq\label{ocan2}
\Omega^{\mathsf{can}} =\frac{1}{8\pi G}\int_{\cN}\ve_{\cN}^{(0)} \Big(\delta\left(\tfrac{1}{2}\Omega \sigma^{ab}\right)\wedge\delta q_{ab}-\delta\mu \wedge \delta\Omega\Big).
\eeq
We can decompose the metric into its unimodular and determinant parts,\footnote{This tensor is degenerate so its determinant vanishes. By det$(\bq)$, we mean the determinant of the non-degenerate part of this tensor, which is co-rank $1$.} such that
\beq
q_{ab}=\Omega\, \bq_{ab}\qquad \text{det}(\bq)=1.
\eeq
Then, the shear can be written as
$
\sigma_a{}^b=\frac{1}{2}\pa_v \bq_{ac}\bq^{cb}
$, where $\bq^{ab}$ satisfies $\bq^{ab} k_b=0$ and $\bq^{ac}\bq_{cb}=q_b{}^a$, with $q_b{}^a$ the horizontal projector.
The Raychaudhuri equation   can then be written, using \eqref{Ray}, as a constraint equation
\beq\label{FinRay}
C= \pa_v^2\Omega-\mu\pa_v\Omega +\Omega\left(\sigma^2 +8\pi G  T^\mathsf{mat}_{vv}\right)=0,
\eeq
where $\sigma^2 := \sigma_{a}{}^{b}\sigma_{b}{}^a$.
We observe that, compared to  \eqref{Ray}, this equation seems more regular in the presence of creases and caustics, where $\Omega\to 0$, provided that $\mu$ has sufficiently regular behaviour. While we have excluded these situations in our analysis, it would be interesting to study the regularity of our constraints further.

The canonical presymplectic $2$-form becomes
\beq
\Omega^{\mathsf{can}}=\frac{1}{8\pi G}\int_{\cN}\ve_{\cN}^{(0)} \Big(\frac{1}{4}\delta\left(\pa_v \bq_{a'b'}\bq^{aa'}\bq^{bb'}\right)\wedge\delta\left(\Omega \bq_{ab}\right)-\delta\mu \wedge \delta\Omega\Big).
\eeq
To invert it and thus derive the kinematic Poisson bracket, it is convenient to  introduce null coframes satisfying $m_a m^a=0=\Bm_b\Bm^b$ and $m_a\Bm^a=1=\Bm_a m^a$, together with $\ell^a m_a=0=\ell^a\Bm_a$ and $m^a k_a=0=\Bm^a k_a$. For the rest of this section indices are raised and lowered using $\bq_{ab}$, so that $m^a=\bq^{ab}m_b$. 
These frames can be used to decompose the metric  and the tensor $\epsilon_a{}^b$ as follows
\bea 
\bq_{ab} &=& m_{a} \Bm_{b}+\Bm_a m_b, \\
\epsilon_{a}{}^b &=& m_{a} \Bm^{b}-\Bm_a m^b.
\eea
The tensor $\epsilon_a{}^b$ satisfies the important properties
\beq
\epsilon_a{}^b\epsilon_b{}^a= 1 \qquad \epsilon_a{}^b m_b = m_a \qquad \epsilon_a{}^b\Bm_b=-\Bm_a.
\eeq
The sphere complex structure will have non-trivial Poisson brackets, and it is given by $J_a{}^b=i\epsilon_a{}^b$.\footnote{In the following, in an abuse of notation, we will refer to $\epsilon_a{}^b$ itself as the complex structure.} The physical results we will derive are real combinations involving this complex structure, which is a useful tensor in the following.

The variation of the coframe is decomposed as
\bea
\delta m_a &=&  \Bm_a\Delta +  m_a\omega\\
\delta \Bm_a &=&  m_a\bD - \Bm_a\omega
\\
\delta m^a&=&- \Bm^a\Delta+ m^a\omega\\
\delta \Bm^a&=&-m^a\bD- \Bm^a\omega.
\eea 
Here we introduced two complex one-forms in field space, $\Delta$ and $\omega$, whose explicit expressions depend on the parameterization chosen. We note that $\omega$ corresponds to a  connection for frame rotations, and that it is pure imaginary, in order to preserve the condition $m_a\Bm^a=1$. 
 
Therefore, we have that 
\bea
\delta \bq_{ab}&=&
2\Bm_a \Bm_b \Delta + 2 m_a m_b\bD,\\
\delta \epsilon_a{}^b&=& 
2 \Bm_a \Bm^b \Delta - 2 m_a m^b\bD.
\eea
We see that both variations are independent of $\omega$.
Further, we note that $m_am^b$ and its complex conjugate form a basis of the space of $2\times 2$ traceless symmetric matrices. This means that the complex structure variation contains the same information as the metric variation
\be 
\delta \epsilon_a{}^b \epsilon_b{}^c=- \epsilon_a{}^b \delta \epsilon_b{}^c = \delta \bq_{ab} \bq^{bc}. 
\ee 

The $v$-derivative of $m$ and $\Bm$ can also be decomposed in a similar fashion
\beqn
\pa_v m_a= \Bm_a\Delta_v+ m_a\omega_v\qquad
\pa_v \Bm_a= m_a\bD_v- \Bm_a\omega_v,
\eeqn
where $\Delta_v$ and $\omega_v$ are two complex scalars given by  $I_{\hat{\pa}_v}\Delta=\Delta_v$, etc. Then
\beq\label{qdq}
\sigma_{a}{}^{b}= \Bm_a \Bm^b\Delta_{v}+   m_a m^b\bD_{v}.
\eeq

A straightforward computation which uses that
\be 
\delta \sigma^{ab} = [(\delta-2\omega)\Delta_v] \Bm^a \Bm^b +
[(\delta+2\omega)\bD_v] m^a m^b -
(\bD_{v}\Delta+\Delta_{v}\bD) \bq^{ab},
\ee 
leads to the final expression 
\beq\label{Ocan2}
\Omega^{\mathsf{can}}
=\frac1{8\pi G} \int_{\cN}\ve_{\cN}^{(0)}
\Big(\delta\Omega\wedge\big(\delta\mu+\bD_{v}\Delta+\Delta_{v}\bD\big)+\Omega \big([(\delta-2\omega)\Delta_v] \wedge \bD+ [(\delta+2\omega)\bD_v]\wedge\Delta\Big)\Big).
\eeq
This is the expression we need to invert, in order to find the kinematic Poisson bracket.

\subsection{Beltrami Parameterization}\label{S42}

To proceed and construct the inverse of the canonical 2-form, it is useful to introduce the Beltrami parameterization \cite{Beltrami}, see also \cite{Reisenberger:2007pq, Reisenberger:2007ku}. This parameterization is a convenient way of describing the degrees of freedom of a $2$-dimensional metric. First of all, given that $\ell=\pa_v$, the condition $\ell^a q_{ab}=0$ reduces to $q_{va}=0=q_{vv}$ and $m_v=0=\Bm_v$. Then, the null coframe can be parameterized as
\beq
m_z =\frac{1}{\sqrt{\beta}}=\Bm_\bz,\qquad m_\bz= \frac{\zeta}{\sqrt{\beta}}, \qquad \Bm_z= \frac{\bzeta}{\sqrt{\beta}},\qquad \beta:=
1-\zeta\bzeta,
\eeq
where we introduced the Beltrami differential $\zeta(x)$ and its complex conjugate $\bzeta(x)$, such that the unimodular metric can be written as (recall $x^a=(v,z,\bz)$)
\beq
\bq_{ab}(x)\rd x^a \rd x^b=2(m_a\rd x^a)(\Bm_b\rd x^b)=2\frac{\vert \rd z+\zeta \rd \bz\vert^2}{\beta}.
\eeq
Examining the variation of the null frame given above leads to the explicit formulae
\beqn
\Delta=\frac{\delta\zeta}{\beta},\qquad \bar\Delta=\frac{\delta\bzeta}{\beta},\qquad \omega=\frac{\zeta\delta\bzeta-\bzeta\delta\zeta}{2\beta}=-\overline{\omega}.
\eeqn
Similarly, the $v$-derivative of $m$ and $\Bm$ gives
\beqn
\Delta_v=\frac{\pa_v\zeta}{\beta},\qquad \bD_v=\frac{\pa_v\bzeta}{\beta},\qquad \omega_v=\frac{\zeta\pa_v\bzeta-\bzeta\pa_v\zeta}{2\beta}=-\overline{\omega}_v,
\eeqn
which again are related to the variations by $I_{\hat{\pa}_v}\Delta=\Delta_v$, etc. 

With the Beltrami differential, we can process \eqref{Ocan2} further. We begin by noting the identities
\beq
&(\delta-2\omega)\Delta_v=\frac{\delta\pa_v \zeta}{\beta}+2\frac{\delta\zeta\bzeta\pa_v\zeta}{\beta^2}=(\pa_v-2\omega_v)\Delta,&\\ &(\delta+2\omega)\bD_v=\frac{\delta\pa_v \bzeta}{\beta}+2\frac{\delta\bzeta\zeta\pa_v\bzeta}{\beta^2}=(\pa_v+2\omega_v)\bD.&
\eeq
Introducing the $\omega$-covariant derivatives
\beq
D_v=\pa_v-2\omega_v,\qquad \oD_v=\pa_v+2\omega_v,
\eeq
we arrive at
\beq
\Omega^{\mathsf{can}}
=\frac1{8\pi G} \int_{\cN}
\ve_{\cN}^{(0)}\Big(\delta\Omega\wedge\big(\delta\mu+\bD_{v}\Delta+\Delta_{v}\bD\big)+\Omega \left(D_v\Delta \wedge \bD+ \oD_v\bD\wedge\Delta\right)\Big).
\eeq
In this section, we will be  interested in the constraints in the bulk of the null hypersurface. Therefore, we assume that the variations of the Beltrami differentials vanish at the corner, such that we can freely integrate by parts in $v$ this expression to  rewrite it as
\beq\label{om1}
\Omega^{\mathsf{can}}
=\frac1{8\pi G}\int_{\cN}
\ve_{\cN}^{(0)}\Big(\delta\Omega\wedge\big(\delta\mu+\bD_{v}\Delta+\Delta_{v}\bD\big)+2D_v(\sqrt{\Omega}\Delta)\wedge\sqrt{\Omega}\bD\Big).
\eeq
In the next subsection, we will derive the kinematic Poisson bracket from this canonical presymplectic structure. Note that we may already expect  some complications, as the spin-0 and spin-2 degrees of freedom are coupled together.

\subsection{Propagators}\label{S43}

To properly construct the inverse of \eqref{om1}, a crucial ingredient is a certain propagator. Denoting $x_1=(v_1,z_1,\bz_1)$ and $x_2=(v_2,z_2,\bz_2)$, this is defined as\footnote{We call these quantities "propagators" as they can be intuitively seen as non-trivial holonomies connecting (propagating) the geometric data between the null times $v_1$ and $v_2$.}
\beq\label{propa}
\bcP_{12}=\frac{1}{\sqrt{\Omega_1\Omega_2}}
e^{-2\int_{v_1}^{v_2}\omega_v \rd v}H(v_1-v_2)\delta^{(2)}(z_1-z_2),
\eeq
where $\Omega_1=\Omega(x_1)$ etc., and $H(v)$ is the odd Heaviside function discussed in Appendix \ref{A1}.
Given that $\overline{\omega}_v=-\omega_v$, it satisfies the property
\beq
\cP_{12}=\overline{\bcP}_{12}=\frac{1}{\sqrt{\Omega_1\Omega_2}}
e^{2\int_{v_1}^{v_2}\omega_v \rd v}H(v_1-v_2)\delta^{(2)}(z_1-z_2)=-\bcP_{21}.
\eeq

This propagator is crucial because it is the covariant antiderivative of the delta function. Indeed, using $\theta_1=\pa_{v_1}\log\Omega_1$, we have
\beq\label{dbpd}
\left(\oD_{v_1}+\frac12 \theta_1\right)\bcP_{12}=\frac{1}{\sqrt{\Omega_1\Omega_2}}
e^{-2\int_{v_1}^{v_2}\omega_v \rd v}\pa_{v_1}H(v_1-v_2)\delta^{(2)}(z_1-z_2)=\frac{\delta^{(3)}(x_1-x_2)}{\Omega_1},
\eeq
where we denote $\delta^{(3)}(x):=\delta(v)\delta^{(2)}(z)$.
Note that this implies
\beq\label{dbpd2}
\left(D_{v_2}+\frac12 \theta_2\right)\bcP_{12}=\frac{1}{\sqrt{\Omega_1\Omega_2}}
e^{-2\int_{v_1}^{v_2}\omega_v \rd v}\pa_{v_2}H(v_1-v_2)\delta^{(2)}(z_1-z_2)=-\frac{\delta^{(3)}(x_1-x_2)}{\Omega_2}.
\eeq

Using this, we can obtain the kinematic Poisson bracket. We relegate details to Appendix \ref{A1}, and show here the final result, which is
\beq
\{\Omega_1,\mu_2\}&=&8\pi G\,\delta^{(3)}(x_1-x_2)\label{omu}\\\label{mumu}
\{\mu_1,\mu_2\}&=&4\pi G\left(\Delta_{v_1}\bD_{v_2}\bcP_{12}+\bD_{v_1}\Delta_{v_2}\cP_{12}\right)\\
\{\mu_1,\zeta_2\}&=&-4\pi G\,\Delta_{v_1} \beta_2\bcP_{12}\\
\{\mu_1,\bzeta_2\}&=&-4\pi G\,\bD_{v_1} \beta_2\cP_{12}\\
\{\zeta_1,\bzeta_2\}&=&4\pi G\,\beta_1\beta_2\cP_{12}\\
\{\bzeta_1,\zeta_2\}&=&4\pi G\,\beta_1\beta_2\bcP_{12}.
\eeq
These are the fundamental building blocks from which we will derive  brackets among  composite functionals. The spin-$2$ brackets are compatible with those found in \cite{Reisenberger:2007ku}. We observe that the field $\mu$ does not commute with the metric determinant or itself. We expect that this feature will have important repercussions on the quantum theory. 

As shown in Appendix \ref{A2}, from here one can compute the bracket of the complex structure
\beq
\{\epsilon^{1 \, b}_{a},\epsilon^{2 \, d}_{c}\}=-16\pi G\,\cP_{12}\Bm_a^1\Bm_1^b m^2_c m_2^d-16\pi G\,\bcP_{12}m^1_a m^b_1\Bm^2_c\Bm_2^d,
\eeq
which shows that there is a non-trivial non-commutative structure for the geometric data.

Furthermore, we also derive in Appendix \ref{A3} the shear tensor bracket
\beq\label{ss}
\{\sigma^{1b}_a,\sigma^{2d}_c\}
&=&4\pi G\Big(\Bm^1_a\Bm_1^b \epsilon^{2d}_c\Delta_{v_2}D_{v_1}\cP_{12}+\epsilon^{1b}_a \Bm^2_c \Bm_2^d\Delta_{v_1}D_{v_2}\bcP_{12}-\bD_{v_1}\Delta_{v_2}\cP_{12}\epsilon^{1b}_a\epsilon^{2d}_c\cr
&&+m^1_a m_1^b\Bm^2_c \Bm_2^d\overline{D}_{v_1}D_{v_2}\bcP_{12}+c.c.\Big),
\eeq
where $c.c.$ denotes complex conjugation. This result is important in the derivation of the constraint algebra, together with
\beq
\{\sigma^{1b}_{a}\sigma^{1a}_{b},\mu_2\}=2\{\Delta_{v_1}\bD_{v_1},\mu_2\}
=-8\pi G\,\Delta_{v_1}\bD_{v_1}\left( \oD_{v_1}\bcP_{12}+D_{v_1}\cP_{12}\right).
\eeq

\subsection{Constraint Algebra}\label{S44}

The Hamiltonian associated to diffeomorphisms along $\ell$ was given above in \eqref{Mf}. We can split it into its spin-0 and spin-2 contributions\footnote{In this section, since we are interested in the constraints, we are not considering corner contributions. Therefore, $M'=M$, but we are working in the framework of subsection \ref{S32}, and thus on the field space spanned by $\fL'$. We also assume that the flux vanish, hence $f|_{\cC}$.}
\beq
M'_f=\frac1{8\pi G} \int_{\cN}\ve^{(0)}_{\cN}\Omega\left( f\sigma_a{}^b\sigma_b{}^a-\theta (\ell+\mu)[f] \right)=M^{(2)}_f+M^{(0)}_f,
\eeq
such that
\beq\label{spin2}
M_f^{(2)}:=\frac1{8\pi G} \int_{\cN}\ve^{(0)}_{\cN}\Omega f\sigma_a{}^b\sigma_b{}^a,
\eeq
and 
\beq\label{spin0}
M_f^{(0)}:=-\frac1{8\pi G} \int_{\cN}\ve^{(0)}_{\cN}\Omega\theta (\ell+\mu)[f] = \frac1{8\pi G} \int_{\cN}\ve^{(0)}_{\cN} (\pa^2_v\Omega-\pa_v\Omega\mu)f.
\eeq

We wish to compute the brackets of these  spin-0 and spin-2 contributions. Instead of simply reporting the final result, it is instructive to present here a detailed derivation. This serves as an illustrative example of the kind of manipulations needed in this section. We start with the spin-2 piece. Using that $\{\Omega_1,\zeta_2\}=0=\{\Omega_1,\bzeta_2\}$, we get
\beq
\{M^{(2)}_f,M^{(2)}_g\}&=&\frac1{(8\pi G)^2}
\int_{{\cN}_1}\ve_{{\cN}_1}^{(0)}\int_{{\cN}_2}\ve_{{\cN}_2}^{(0)}\, f_1\Omega_1g_2\Omega_2\{\sigma^{1a}_b\sigma^{1b}_a,\sigma^{2c}_d\sigma^{2d}_c\}\\
&=&\frac1{(4\pi G)^2}\int_{{\cN}_1}\ve_{{\cN}_1}^{(0)}\int_{{\cN}_2}\ve_{{\cN}_2}^{(0)}\, f_1\Omega_1g_2\Omega_2\sigma_b^{1a} \{\sigma^{1b}_a,\sigma^{2d}_c\} \sigma_d^{2c}.
\eeq
Inserting \eqref{ss}, and using that $\sigma_a{}^b\epsilon_b{}^a=0$, we process this expression into the form
\beq
\{M^{(2)}_f,M^{(2)}_g\}&=&\frac1{4\pi G}\int_{{\cN}_1}\ve_{{\cN}_1}^{(0)}\int_{{\cN}_2}\ve_{{\cN}_2}^{(0)}\, f_1\Omega_1g_2\Omega_2\sigma_b^{1a} \Big(m^1_a m_1^b\Bm^2_c \Bm_2^d\overline{D}_{v_1}D_{v_2}\bcP_{12}\cr
&&+\Bm^1_a \Bm_1^bm^2_c m_2^dD_{v_1}\oD_{v_2}\cP_{12}\Big) \sigma_d^{2c}\\
&=&\frac{1}{4\pi G}\int_{{\cN}_1}\ve_{{\cN}_1}^{(0)}\int_{{\cN}_2}\ve_{{\cN}_2}^{(0)}\, f_1\Omega_1g_2\Omega_2\Big(\Delta_{v_1}\bD_{v_2}\overline{D}_{v_1}D_{v_2}\bcP_{12}+\bD_{v_1}\Delta_{v_2}D_{v_1}\oD_{v_2}\cP_{12}\Big) .
\eeq
Then, using \eqref{dbpd}, \eqref{dbpd2}, and their conjugates, we obtain
\beq
\{M^{(2)}_f,M^{(2)}_g\}&=&\frac1{8\pi G}\int_{\cN}\ve_{{\cN}}^{(0)}\,\Omega\Big(D_v(f\Delta_v)g\bD_v-f\Delta_v\oD_v(g\bD_v)+\oD_v(f\bD_v)g\Delta_v-f\bD_vD_v(g\Delta_v)\Big)\cr
&&+\frac1{16\pi G}\int_{{\cN}_1}\ve_{{\cN}_1}^{(0)}\int_{{\cN}_2}\ve_{{\cN}_2}^{(0)}\, f_1\pa_{v_1}\Omega_1g_2\pa_{v_2}\Omega_2\left(\Delta_{v_1}\bD_{v_2}\bcP_{12}+\bD_{v_1}\Delta_{v_2}\cP_{12}\right)\\
&=&-\frac1{4\pi G}\int_{\cN}\ve_{{\cN}}^{(0)}\,\Omega(f\pa_v g-g\pa_v f) \Delta_v\bD_v\cr
&&+\frac1{16\pi G}\int_{{\cN}_1}\ve_{{\cN}_1}^{(0)}\int_{{\cN}_2}\ve_{{\cN}_2}^{(0)}\, f_1\pa_{v_1}\Omega_1g_2\pa_{v_2}\Omega_2\left(\Delta_{v_1}\bD_{v_2}\bcP_{12}+\bD_{v_1}\Delta_{v_2}\cP_{12}\right)\\
&=&-M^{(2)}_{f\pa_v g-g\pa_v f}+\int_{{\cN}_1}\ve_{{\cN}_1}^{(0)}\int_{{\cN}_2}\ve_{{\cN}_2}^{(0)}\, \frac{f_1\pa_{v_1}\Omega_1g_2\pa_{v_2}\Omega_2}{16\pi G}\left(\Delta_{v_1}\bD_{v_2}\bcP_{12}+\bD_{v_1}\Delta_{v_2}\cP_{12}\right),
\eeq
where we also used $\sigma_a{}^b\sigma_b{}^a=2\Delta_v \bD_v$. 

{\it We conclude that the spin-$2$ algebra does not close.} Comparing the last term with \eqref{mumu}, we rewrite the result as 
\beq
\{M^{(2)}_f,M^{(2)}_g\}=-M^{(2)}_{f\pa_v g-g\pa_v f}+\frac1{(8\pi G)^2}\int_{{\cN}_1}\ve_{{\cN}_1}^{(0)}\int_{{\cN}_2}\ve_{{\cN}_2}^{(0)}\, f_1\pa_{v_1}\Omega_1g_2\pa_{v_2}\Omega_2\{\mu_1,\mu_2\}.
\eeq
This proves that the non-closure of the spin-$2$ sector is entirely due to the non-commutativity of the field $\mu$, that one can trace back to the coupling of the spin-$0$ and spin-$2$ fields in the presymplectic form. 

Similarly, for the spin-$0$ sector, we have
\beq
\{M_f^{(0)},M_g^{(0)}\}
&=&\frac1{(8\pi G)^2} \int_{{\cN}_1}\ve_{{\cN}_1}^{(0)}\int_{{\cN}_2}\ve_{{\cN}_2}^{(0)}\,  f_1g_2\{\pa^2_{v_1}\Omega_1-\pa_{v_1}\Omega_1\mu_1,\pa^2_{v_2}\Omega_2-\pa_{v_2}\Omega_2\mu_2\}\\
&=& \int_{\cN}\frac{\ve_{{\cN}}^{(0)}\pa_{v}\Omega}{8\pi G}\left(f\pa^2_{v}g-g\pa^2_{v}f\right)+ \int_{{\cN}_1}\ve_{{\cN}_1}^{(0)}\int_{{\cN}_2}\ve_{{\cN}_2}^{(0)}\, \frac{f_1g_2}{(8\pi G)^2}\{\pa_{v_1}\Omega_1\mu_1,\pa_{v_2}\Omega_2\mu_2\}\\
&=&-\frac1{8\pi G} \int_{\cN}\ve_{{\cN}}^{(0)}\left(f\pa_{v}g-g\pa_{v}f\right)\pa^2_{v}\Omega+\frac1{(8\pi G)^2} \int_{{\cN}_1}\ve_{{\cN}_1}^{(0)}\int_{{\cN}_2}\ve_{{\cN}_2}^{(0)}\, f_1g_2\cr
&&\times\Big(\mu_1\pa_{v_2}\Omega_2\{\pa_{v_1}\Omega_1,\mu_2\}+\pa_{v_1}\Omega_1\{\mu_1,\pa_{v_2}\Omega_2\}\mu_2+\pa_{v_1}\Omega_1\pa_{v_2}\Omega_2\{\mu_1,\mu_2\}\Big)\\
&=&-\frac1{8\pi G} \int_{\cN}\ve_{{\cN}}^{(0)}\left(f\pa_{v}g-g\pa_{v}f\right)\left(\pa^2_{v}\Omega-\pa_v\Omega \mu\right)\nonumber\\
&&+\frac1{(8\pi G)^2} \int_{{\cN}_1}\ve_{{\cN}_1}^{(0)}\int_{{\cN}_2}\ve_{{\cN}_2}^{(0)}\, f_1g_2\pa_{v_1}\Omega_1\pa_{v_2}\Omega_2\{\mu_1,\mu_2\}\\
&=&-M^{(0)}_{f\pa_v g-g\pa_v f}+\frac1{(8\pi G)^2} \int_{{\cN}_1}\ve_{{\cN}_1}^{(0)}\int_{{\cN}_2}\ve_{{\cN}_2}^{(0)}\, f_1g_2\pa_{v_1}\Omega_1\pa_{v_2}\Omega_2\{\mu_1,\mu_2\}.
\eeq
The same bi-local contribution seen in the spin-$2$ sector appears here as well, and thus also the spin-$0$ sector does not close. 

As expected, but nevertheless quite remarkably, the mixed bracket contains exactly this extra term. Utilizing similar manipulations as those used for the spin-$0$ algebra, one arrives at\footnote{By symmetry, one can also show $\{M^{(0)}_f,M^{(2)}_g\}=\{M^{(2)}_f,M^{(0)}_g\}$.}
\beq
\{M^{(2)}_f,M^{(0)}_g\}=-\frac1{(8\pi G)^2} \int_{{\cN}_1}\ve_{{\cN}_1}^{(0)}\int_{{\cN}_2}\ve_{{\cN}_2}^{(0)}\, f_1 g_2\pa_{v_2}\Omega_2\pa_{v_1}\Omega_1\{\mu_1,\mu_2\},
\eeq
such that the algebra of the full constraints properly closes
\beq
\{M'_f,M'_g\}=\{M^{(2)}_f+M^{(0)}_f,M^{(2)}_g+M^{(0)}_g\}
=-M^{(2)}_{(f\pa_v g-g\pa_v f)}-M^{(0)}_{(f\pa_v g-g\pa_v f)}=-M'_{(f\pa_v g-g\pa_v f)}.
\eeq
So the split into the spin-$0$ and spin-$2$ part is a delicate procedure. We expect that this split can be consistently done perturbatively in $G$. We are thus unveiling here a non-perturbative effect mixing graviton propagation and the underlying geometry of null hypersurfaces. A full quantum theory would  therefore be complete only when keeping the spin-$0$ geometric data. Indeed, one should think of the spin-$0$ sector (as well as the disregarded spin-$1$ sector) as completing the extension of the phase space of the theory such that a proper representation of the constraint algebra is obtained, and the constraints act canonically on the fields, as we now show.

Indeed, before concluding this section we display the action of the spin-$2$ and spin-$0$ Hamiltonians on our fields. A long yet straightforward computation gives
\beq
\{M^{(2)}_f,\mu(x_2)\}=\frac1{8\pi G}\int_{{\cN}_1}\ve^{(0)}_{{\cN}_1}f(x_1)\pa_{v_1}\Omega(x_1)\{\mu(x_1),\mu(x_2)\},
\eeq
while
\beq
\{M^{(0)}_f,\mu(x_2)\}=\pa^2_{v_2}f(x_2)+\pa_{v_2}(\mu(x_2)f(x_2))-\frac1{8\pi G}\int_{{\cN}_1}\ve^{(0)}_{{\cN}_1}f(x_1)\pa_{v_1}\Omega(x_1)\{\mu(x_1),\mu(x_2)\},
\eeq
such that
\beq
\{M'_f,\mu(x_2)\}=\{M^{(0)}_f+M^{(2)}_f,\mu(x_2)\}
=\pa^2_{v_2}f(x_2)+\pa_{v_2}(\mu(x_2)f(x_2)).
\eeq
Not only this is an important consistency check (see \eqref{mutr}), it also shows how both the spin-$0$ and spin-$2$ Hamiltonians are needed to recover the correct $\mu$-transformation. 

This is not the case for $\Omega$, on which only the spin-$0$ Hamiltonian acts non-trivially
\beq
\{M^{(2)}_f,\Omega(x_2)\}&=&0,\\
\{M^{(0)}_f,\Omega(x_2)\}&=&f(x_2)\pa_{v_2}\Omega(x_2),
\eeq
which again shows that the constraint acts canonically \eqref{omtr}.

\section{Time and Boost Operator}\label{S5}

In the preceding discussion, we have used a kinematical coordinate $v$ to parameterize null curves along the Carroll structure, and we have seen that the kinematic Poisson brackets involve the spin-0 fields in a non-trivial way. Their inclusion has the important effect of giving a proper representation of the constraint algebra. In this section, we show that time can be promoted to a dynamical variable conjugate to the Raychaudhuri constraint. We focus on a particular choice of time, called the dressing time,  where the time variable is constructed as the holonomy of the surface tension. In this frame the spin-$0$ data are relegated to the corner. Furthermore, the constraint can be solved non-perturbatively on sections of the null surface where the expansion doesn't vanish.

\subsection{Time  Operator}

We begin by considering a diffeomorphism $V:v\to V(v,\sigma)$, combined with a local boost such that $\ell$ remains equal to $\pa_v$. 
The phase space variables
$(\Omega , \mu , q_{ab}, \sigma_{ab})$ can be written as functions of $V$ in terms of their pullbacks, 
 which we formalize as follows, 
\bea
&\Omega= 
\tilde\Omega \circ V, \qquad
q_{ab}= 
\tilde q_{ab}\circ V, \qquad 
\sigma_{ab}= 
\pa_v V(\tilde\sigma_{ab}\circ V),\qquad \mu= 
\pa_vV (\tilde{\mu}\circ V) + \frac{\pa_v^2 V}{\pa_v V}.&
\label{finitetransmu}
\eea
By the composition of maps we mean that $\Omega(v)=\tilde\Omega\circ V(v)=\tilde\Omega(V(v))$, etc.  We note that $\mu$ transforms non-linearly: this is due to the fact that it is a boost connection, and as noted above, we are considering a combined  diffeomorphism and internal boost.\footnote{In \eqref{finitetransmu}, we are giving a finite transformation.  Thus the given non-linear shift in the transformation of $\mu$  corresponds to a finite transformation of the form $\lambda_V^{-1}\pa_v\lambda_V$, with $\lambda_V=\pa_vV$. This is an instance of the combined $\ell$-diffeomorphism/boost whose infinitesimal form was given in \eqref{combinedsupertrans}. } Its transformation ensures that if $O$ is a tensor of boost weight $s$ such that $O=
(\pa_vV)^s (\tilde{O}\circ V)$ then 
\be 
(\pa_v - s \mu) O= (\pa_vV)^{s+1} [(\pa_V -s \tilde{\mu})\tilde{O}]\circ V, 
\ee is a tensor of boost weight $s+1$.  In this equation we use that everything is happening locally in $\sigma$. For instance, for a boost weight-$0$ tensor $O$, we have that $O(v,\sigma)=\tilde{O}(V(v,\sigma), \sigma)$ is simply denoted $\tilde{O}\circ V$ and by $\pa_V$ we mean the derivative keeping $\sigma$ fixed.\footnote{Of course when we change frame we also have to change the tangential derivative 
$ \pa_A O = (\tilde\pa_A \tilde{O})\circ V$, where 
$\tilde\pa_A=\pa_A - \pa_AV \pa_v$ is the tangential derivative along $V=$cst.} So in the following  we regard $V=V(v)$ as a shortcut notation.
 We note that the Raychaudhuri constraint $C= \pa_v^2\Omega-\mu\pa_v\Omega + \Omega (\sigma_{a}{}^{b}\sigma_{b}{}^{a}+8\pi G\,T^{\mathsf{mat}}_{vv})$ is of weight two and thus satisfies
\beq\label{cct}
C=(\pa_vV)^2\tilde C\circ V
\eeq
where $\tilde C= \pa_V^2\tilde \Omega-\tilde \mu\pa_V\tilde \Omega + \tilde \Omega (\tilde \sigma_{a}{}^{b}\tilde \sigma_{b}{}^{a}+8\pi G\,\tilde T^{\mathsf{mat}}_{VV})$.

The non-linear transformation of $\mu$  means that  it is possible to choose the time variable $V$ such that $\tilde\mu=0$, or in other words
\be \label{introV}
\mu =\frac{\pa_v^2 V}{\pa_v V}.
\ee 
This should be read as trading the dynamical variable $\mu$ for $V$.
The quantity $V$ can be thought of as a  "rest frame" time for time itself where the boost connection vanishes. In the following we refer to $V$ satisfying \eqref{introV} as the \emph{dressing time}.
It is important to appreciate that the choice \eqref{introV} corresponds to employing a dynamical field variable as a clock.

This relation defines $V$ only up to an affine transformation
$V\to AV+B$, which means that there are residual symmetries preserving $\tilde\mu=0$. These are the BMSW transformations discussed in Section \ref{S33}. Eq. \eqref{introV} is easily integrated in terms of two pieces of data, the values of $V(v,\sigma)$ and $\pa_vV(v,\sigma)$ at particular values of $v$, which we write as $a,b$. The general solution is
\be 
V_{(a,b)}(v, \sigma)=
V(a,\sigma)+\pa_vV(b,\sigma)
\int_a^v \rd v' \exp \left(\int_b^{v'}\rd v" \mu(v",\sigma) \right).
\ee 
We note that whereas $\mu$ transforms as a connection under the internal boost symmetry, $V$ transforms as a Goldstone field.  The original  field space variables are dressed version of the tilded variables $\tilde\sigma_{ab}$, etc., which are therefore invariant under diffeomorphism.
From this expression we see that $\pa_v V = \exp \left(\int_b^{v} \mu(v',\sigma) \rd v'\right)$ 
is simply a Wilson line for the boost connection. It is invertible unless $\mu=\pm \infty$. Suppose we extend our construction through a caustic, where $\theta\to \pm\infty$. If $\kappa$ remains finite, one has that $\mu\to \pm\infty$, which is indeed this singular point in the Jacobian.

An analogous transformation is routinely applied in the case where the expansion vanishes such as on a Killing horizon. In this case the \ctime $V$ is equal to the affine time $\lambda$, because $\mu=\kappa$, and thus  
$\kappa = \pa_v^2 \lambda/\pa_v\lambda$, which using notation similar to the above corresponds to $\tilde\kappa=\kappa_{\mathsf{aff}}=0$. Conversely, for an expanding surface, since $\mu= \kappa + \frac{\theta}{2}$, the surface tension differs from the inaffinity and the two times are different.
We can express that $V= V_{\mathsf{aff}} \circ \lambda$ where $ V_{\mathsf{aff}}(\lambda,\sigma)$ is the relative time given by\footnote{The square root appears in $4$ bulk dimensions. In spacetime dimension $d$ we have $ V_{\mathsf{aff}}(\lambda) = A \int_b^{\lambda} \rd \lambda'[\Omega_{\mathsf{aff}}(\lambda')]^{\frac{d-3}{d-2}}$ instead.} 
\be 
 V_{\mathsf{aff}}(\lambda,\sigma) = A \int_b^\lambda   \, \rd \lambda' \sqrt{\Omega_\mathsf{aff}(\lambda',\sigma)}
\ee
where $A,b$ are constants and  we denote $\Omega= \Omega_\mathsf{aff}\circ \lambda$. This follows simply from the chain rule,\footnote{Explicitly, $\pa_v(V_{\mathsf{aff}}\circ \lambda) = \pa_v\lambda (\pa_\lambda V_{\mathsf{aff}}) \circ \lambda $.} which gives
\beq
\frac{\pa_v^2V}{\pa_vV}=\frac{\pa_v^2\lambda}{\pa_v\lambda}+\pa_v\lambda\left( \frac{\pa_\lambda^2V_{\mathsf{aff}}}{\pa_\lambda V_{\mathsf{aff}}}\right) \circ \lambda
\eeq
together with the fact that $\mu_{\mathsf{aff}}=\frac{\pa_\lambda^2V_{\mathsf{aff}}}{\pa_\lambda V_{\mathsf{aff}}}=\frac12{\theta_{\mathsf{aff}}}$. 

We conclude this discussion by noticing that the Raychaudhuri equation can be solved non-perturbatively. Solving Raychaudhuri  by choosing the area element $\Omega:v\to \Omega(v)$ as a time-coordinate. This is possible if one assumes that  the expansion $\pa_v\Omega$ does not vanish.\footnote{When the expansion vanishes, solving Raychaudhuri imposes that  gravitational and matter states are in their vacuum, if we assume positivity of energy. } We can express this {\it areal clock} relative to the dressing clock as
\be
V= \bar{V}\circ \Omega,
\ee
such that
\beq
\mu=\frac{\pa_v^2V}{\pa_vV}=\frac{\pa_v^2\Omega}{\pa_v\Omega}+\pa_v\Omega\left(\frac{\pa_\Omega^2\bar V}{\pa_\Omega\bar V}\right)\circ\Omega.
\eeq
We can then add and subtract the Raychaudhuri constraint to write
\beq
\mu=\frac{\pa_v^2\Omega}{\pa_v\Omega}+\pa_v\Omega\left(\frac{\pa_\Omega^2\bar V}{\pa_\Omega\bar V}\right)\circ\Omega=\mu+\frac{C}{\pa_v\Omega}-\frac{\Omega(\sigma^2+8\pi GT^{\mathsf{mat}}_{vv})}{\pa_v\Omega}+\pa_v\Omega\left(\frac{\pa_\Omega^2\bar{V}}{\pa_\Omega \bar{V}}\right)\circ\Omega.
\eeq

Denoting the shear and the matter stress tensor in the areal frame  
respectively by $(\bar{\sigma}, \bar{T}^{\mathsf{mat}})$ where $\sigma =\pa_v  \Omega (\bar\sigma \circ \Omega)$ 
and $T^{\mathsf{mat}}_{vv}= (\pa_v\Omega)^2 \bar{T}^{\mathsf{mat}}_{\Omega\Omega}\circ \Omega$, we see that on-shell this reduces to 
\beq
\left(-\Omega(\bar\sigma^2+8\pi G\bar{T}^{\mathsf{mat}}_{\Omega\Omega})+\frac{\pa_\Omega^2\bar{V}}{\pa_\Omega \bar{V}}\right)\circ\Omega=0,
\eeq
which integrates to
\beq\label{solRaych}
\pa_\Omega \bar{V}=
\exp\int^\Omega_b \D\Omega'\,\Omega'(\bar\sigma^2+8\pi GT^{\mathsf{mat}}_{\Omega\Omega})(\Omega').
\eeq
This is a non-perturbative solution of the Raychaudhuri equation. We note that in the absence of shear, the exponential's argument is exactly the boost Hamiltonian, where the right-hand side is the matter ANEC operator in the areal time. 

\subsection{Presymplectic Structure in Dressing Time}\label{S52}

The significance of the dressing time can be seen more clearly by considering the presymplectic structure. We are interested here in the general case, where both gravity and matter are present. 
To evaluate the presymplectic structure  we compute the variation of the phase space variables in the dressing frame, where $
\tilde
\mu=0$,
\bea
\delta \mu 
&=&  \frac{\pa_v^2 \delta V}{\pa_v V}- \mu\frac{ \pa_v \delta V}{\pa_v V}  =
\pa_v \left(\frac{\pa_v \delta V}{\pa_v V}\right) ,\label{dmu}\\ 
\delta \Omega 
&=& (\delta\tilde\Omega) \circ V + (\pa_V\tilde\Omega \circ V) \delta V, \label{dom}\\ 
\delta q_{ab}
&=& (\delta \tilde{q}_{ab})\circ V +  (\pa_V\tilde q_{ab}\circ V) \delta V  ,\\ 
\delta \sigma_{ab}
&=& (\delta \tilde\sigma_{ab}\circ V)\pa_v V  + \pa_v ((\tilde\sigma_{ab} \circ V) \delta V ).
\eea

As a simple model of matter, we consider a real scalar field 
with  presymplectic structure on $\cN$
\beq\label{omma}
\Omega^{\mathsf{mat}} = \int_{\cN}\ve_{\cN}^{(0)}\delta(\Omega\pa_v \varphi) \wedge \delta\varphi,
\eeq and  energy-momentum $T^\mathsf{mat}_{vv}= (\pa_v \varphi)^2$. The discussion could be easily extended to an arbitrary matter sector. The matter fields can be expressed with respect to the dressing clock in a similar fashion to the above,
\be \label{phi}
\varphi=
\tilde\varphi\circ V, \qquad 
\pa_v \varphi = 
\pa_v V (\pa_V\tilde\varphi)\circ V.
\ee

We define the total presymplectic structure 
\beq
\Omega^{\mathsf{tot}}=\Omega^{\mathsf{can}}+\Omega^{\mathsf{mat}},
\eeq
 where $\Omega^{\mathsf{can}}$ is the canonical $2$-form \eqref{2fo}.
From the transformations listed above we can establish that 
it
can be rearranged 
as a sum of a bulk and a corner term, in the dressing frame. We relegate details of this computation to Appendix \ref{appendixclocksymp}, and display here the final result. The bulk term is given by 
\beq
\Omega_{\cN}^{\mathsf{tot}} &=& \int_{\cN}\ve_{\cN}^{(0)}\pa_vV \Big(\frac{1}{16\pi G}\delta (\tilde\Omega \tilde \sigma^{ab})\wedge\delta \tilde{q}_{ab} +\delta\left(\tilde\Omega \pa_V \tilde\varphi\right) \wedge \delta \tilde\varphi
\Big)\circ V\nonumber \\
&&+\frac{1}{8\pi G}\int_{\cN}  \ve_{\cN}^{(0)}  \delta \left(\pa_vV[\tilde{C}\circ V]  \right) \wedge \delta V,\label{bulkom}
\eeq
where 
\beq\label{rmu}
\tilde{C}=\pa_V^2\tilde\Omega+ \tilde\Omega\left( \tilde\sigma^2+8\pi G\,(\pa_V \tilde\varphi)^2\right)
\eeq
is the total Raychaudhuri constraint in the dressing frame. 
We see as expected that the spin-$0$ term has disappeared from the bulk presymplectic form and has been replaced by the pairing between the dressing time and the constraint.

On the other hand, the corner term is given by 
\bea\label{Cornersymp}
\Omega_{\cC}^{\mathsf{tot}} &=&
-\frac1{8\pi G} \int_{\cC}\ve^{(0)}_{\cC} 
\left(
\frac{\pa_v \delta  V }{\pa_v V} \wedge    \delta \Omega - \delta V\wedge  \delta \left(\frac{\pa_v \Omega}{\pa_v V}\right)\right)\nonumber\\
& & +
\int_{\cC}\ve^{(0)}_{\cC} 
  \delta V \wedge 
\left( \frac{\tilde{\Omega}}{16\pi G} \tilde\sigma^{ab}\delta \tilde{q}_{ab} + \tilde{\Omega}\pa_V\tilde\varphi \delta\tilde\varphi  \right)\circ V.
\eea
The terms in the second line  come from  spin-$2$ and  matter fields. They are the flux terms $
 \int_{\cC} \delta V \wedge  \iota_{\pa_v}\tilde\theta^\mathsf{tot}\circ V$, with
$\tilde\theta^{\mathsf{tot}}=\tilde\theta^{(2)}+\tilde\theta^{\mathsf{mat}}=\ve^{(0)}_{\cN}(\frac{\tilde\Omega}{16\pi G}\tilde\sigma^{ab}\delta\tilde q_{ab}+\tilde{\Omega}\pa_V\tilde\varphi \delta\tilde\varphi)$. These are the flux terms introduced in \eqref{fff}, ${\cal F}_f=\int_{\cC}f\iota_{\pa_v}\theta^\mathsf{tot}$.

It is instructive for the computation of the charge to decompose this off-shell symplectic structure in a sum of symplectic terms attached to the spin-$0$, spin-$2$ and matter contributions. There is a bulk split
\be
\Omega^{\mathsf{tot}}_{\cN}= \Omega^{(0)}_{\cN}+\Omega^{(2)}_{\cN}+\Omega^{\mathsf{mat}}_{\cN},
\ee 
and a corner split
\be
\Omega^{\mathsf{tot}}_{\cC}= \Omega^{(0)}_{\cC}+\Omega^{(2)}_{\cC}+\Omega^{\mathsf{mat}}_{\cC}.
\ee 
Splitting likewise the Raychaudhuri constraint $\tilde C=\tilde C^{(0)}+\tilde C^{(2)}+\tilde C^{\mathsf{mat}}$, with
\be
\tilde C^{(0)}=\pa_V^2\tilde \Omega,\qquad \tilde C^{(2)}=\tilde \Omega\,\tilde\sigma^2,\qquad \tilde C^{\mathsf{mat}}=8\pi G\,\tilde\Omega (\pa_V \tilde\varphi)^2,
\ee 
we gather
\beq
\Omega^{(0)}_{\cN}&=&\frac{1}{8\pi G}\int_{\cN}  \ve_{\cN}^{(0)}  \delta \left(\pa_vV\tilde{C}^{(0)}\circ V  \right) \wedge \delta V\\
\Omega^{(2)}_{\cN}&=&\frac{1}{8\pi G}\int_{\cN}  \ve_{\cN}^{(0)}  \delta \left(\pa_vV\tilde{C}^{(2)}\circ V  \right) \wedge \delta V+\int_{\cN}\pa_vV \delta \tilde\theta^{(2)} \circ V\\
\Omega^{\mathsf{mat}}_{\cN}&=&\frac{1}{8\pi G}\int_{\cN}  \ve_{\cN}^{(0)}  \delta \left(\pa_vV\tilde{C}^{\mathsf{mat}}\circ V  \right) \wedge \delta V+\int_{\cN}\pa_vV \delta\tilde\theta^{\mathsf{mat}} 
\circ V.
\eeq
We thus see that the bulk part of the presymplectic structure is the same for the spin-$2$ and matter sectors, while the spin-$0$ sector is different. This is also the case for the corner part of the presymplectic structure
\beq
\Omega^{(0)}_{\cC}&=&-\frac1{8\pi G} \int_{\cC}\ve^{(0)}_{\cC} 
\left(
\frac{\pa_v \delta  V }{\pa_v V} \wedge    \delta \Omega - \delta V\wedge  \delta \left(\frac{\pa_v \Omega}{\pa_v V}\right)\right)\label{oc0}\\
\Omega^{(2)}_{\cC}&=& \int_{\cC} \delta V \wedge  \iota_{\pa_v}\tilde\theta^{(2)}\circ V\label{os2c}\\
\Omega^{\mathsf{mat}}_{\cC}&=& \int_{\cC} \delta V \wedge  \iota_{\pa_v}\tilde\theta^\mathsf{mat}\circ V.\label{omatc}
\eeq
This is the main result of this section. In the dressing time, the presymplectic structure in the spin-$2$ and matter sectors displays a pattern where the corner contribution is given entirely by the canonical flux. As already commented upon, the spin-$0$ sector is peculiar: it has no bulk  potential, $\tilde\theta^{(0)}=0$, and on the corner it gives rise to the charge. The dressing time is therefore guaranteeing a split between the spin-$0$ sector and the rest.

In an arbitrary frame, the spin-$0$ contributes also in the bulk. Indeed, one can just keep all terms containing $\tilde\mu$ in the previous computation. One finds (see appendix \ref{appendixclocksymp}), apart from the additional term involving $\tilde\mu$ appearing in the Raychaudhuri constraint $\tilde C$ and the contribution to the canonical flux $\int_{\cC}\delta V\wedge \iota_{\pa_v}\tilde\theta^{(0)}\circ V$ with $\tilde\theta^{(0)}=-\frac1{8\pi G}\ve^{(0)}_{\cN} \tilde\mu \delta\tilde\Omega $,
 simply an additional term of the form
\be 
-\frac{1}{8\pi G}\int_{\cN}\varepsilon_{\cN}^{(0)} \pa_v V   (\delta \tilde{\mu} \wedge \delta \tilde{\Omega})\circ V.
\ee 
As announced, this means that the symplectic pair $(\tilde\mu,\tilde\Omega)$ remains dynamical in the bulk in an arbitrary frame. 
Thus, to reiterate, the dressing time is special in that it is the unique choice of clock for which the spin-0 fields are removed from the bulk symplectic structure, with the clock variable $V$ appearing conjugate to the constraint $\tilde C$. 
This result has far reaching consequences: it implies, as we are about to see, that the canonical boost charge in the dressing time is monotonic.

\subsection{Boost Operator}\label{sec53}

Now that we have introduced the time $V$ and the corresponding dressed fields $(\tilde{q}_{ab},\tilde{\sigma}_{ab})$, it is interesting to redo the canonical analysis in their terms.
First, the  dressing time transforms under infinitesimal diffeomorphisms simply as 
 \be \label{Vt}
 \fL'_{\hat f} V = f\pa_v V.
 \ee 
 This transformation implies \eqref{mutr}. 
 Then, the dressed variables are by construction gauge invariant\footnote{In an arbitrary frame, one also has $\fL'_{\hat f}\tilde\mu=0$.}
 \be \label{qT}
 \fL'_{\hat f} \tilde\Omega = 0,\qquad \fL'_{\hat f} \tilde{q}_{ab} = 0, \qquad \fL'_{\hat f} \tilde{\sigma}_{ab} = 0, \qquad \fL'_{\hat f} \tilde\varphi=0.
 \ee 
 The proof of this follows from the fact that, on one hand, we know (see \eqref{trs}) that $\fL'_{\hat f} q_{ab}=f \pa_v q_{ab}$, while on the other hand we have 
 \be
 \fL'_{\hat f} q_{ab}= \fL'_{\hat f} (\tilde{q}_{ab} \circ V) = (\fL'_{\hat f}\tilde{q}_{ab}) \circ V+ (\pa_V\tilde{q}_{ab}) \circ V\, \fL'_{\hat f}V.
 \ee 
Under the transformations (\ref{Vt},\ref{qT}), this expression gives $f \pa_v (\tilde{q}_{ab}\circ V)$. A similar analysis can be done for the other fields. 

One can  now evaluate the charge. To do so, we use that $\fL'_{\hat f}[\pa_v V \tilde{C}\circ V]= \pa_v [f\pa_v V \tilde C\circ V]$. Let us show the result sector by sector, starting from the bulk contributions
\beq
I'_{\hat f}\Omega_{\cN}^{(0)}&=&- \frac1{8\pi G}\delta \left(\int_{\cN}\ve^{(0)}_{\cN}f(\pa_v V)^2 \tilde C^{(0)}\circ V\right)
+\frac1{8\pi G}\int_{\cC} \ve_{\cC}^{(0)}f\pa_v V \delta V\tilde C^{(0)}\circ V\\
I'_{\hat f}\Omega_{\cN}^{(2)}&=&- \frac1{8\pi G}\delta \left(\int_{\cN}\ve^{(0)}_{\cN}f(\pa_v V)^2 \tilde C^{(2)}\circ V\right)
+\frac1{8\pi G}\int_{\cC} \ve_{\cC}^{(0)}f\pa_v V \delta V\tilde C^{(2)}\circ V\\
I'_{\hat f}\Omega_{\cN}^{\mathsf{mat}}&=&- \frac1{8\pi G}\delta \left(\int_{\cN}\ve^{(0)}_{\cN}f(\pa_v V)^2 \tilde C^{\mathsf{mat}}\circ V\right)
+\frac1{8\pi G}\int_{\cC} \ve_{\cC}^{(0)}f\pa_v V \delta V\tilde C^{\mathsf{mat}}\circ V,
\eeq
and the corner contributions (we use \eqref{introV}, and process the spin-$0$ part)
\beq
I'_{\hat f}\Omega_{\cC}^{(0)}&=&-\frac1{8\pi G} \int_{\cC}\ve^{(0)}_{\cC} 
f\left(\mu\delta\Omega+ \pa_v V[\pa^2_V\tilde\Omega\circ V]\delta V\right)-\frac1{8\pi G} \delta\left(\int_{\cC}\ve^{(0)}_{\cC} (
\Omega\pa_v f-f\pa_v\Omega)\right)\\
I'_{\hat f}\Omega_{\cC}^{(2)}&=&\int_{\cC} f\pa_v V   [\iota_{\pa_v}\tilde\theta^{(2)}\circ V]\\
I'_{\hat f}\Omega_{\cC}^{\mathsf{mat}}&=&\int_{\cC} f\pa_v V   [\iota_{\pa_v}\tilde\theta^\mathsf{mat}\circ V].
\eeq
For the spin-$2$ and matter parts, an explicit computation shows that the corner contributions coming from the bulk constraint combine with the dressing-time flux to produce the full canonical flux
\beq
&\pa_v V\left(\frac1{8\pi G}\ve_{\cC}^{(0)}[\tilde C^{(2)}\circ V] \delta V +   \iota_{\pa_v}\tilde\theta^{(2)}\circ V\right)=  \iota_{\pa_v}\theta^{(2)},&\\
&\pa_v V\left(\frac1{8\pi G}\ve_{\cC}^{(0)} [\tilde C^{\mathsf{mat}}\circ V]\delta V+ \iota_{\pa_v}\tilde\theta^\mathsf{mat}\circ V\right)=\iota_{\pa_v}\theta^{\mathsf{mat}}.&
\eeq
On the other hand, the same procedure for  spin-$0$ gives  rise to the canonical charge:
\beq
\frac1{8\pi G}\int_{\cC} \ve_{\cC}^{(0)}f\pa_v V [\tilde C^{(0)}\circ V]\delta V+I'_{\hat f}\Omega_{\cC}^{(0)}=\int_{\cC} f  \iota_{\pa_v}\theta^{(0)}-\frac1{8\pi G} \delta\left(\int_{\cC}\ve^{(0)}_{\cC}(\Omega\pa_v f- f\pa_v\Omega)\right).
\eeq
In these expressions, we have introduced the symplectic potentials $\theta^{(0)}=-\frac1{8\pi G}\ve^{(0)}_{\cN}\mu\delta \Omega$, $\theta^{(2)}=\ve^{(0)}_{\cN}\frac{\Omega}{16\pi G}\sigma^{ab}\delta q_{ab}$, and $\theta^{\mathsf{mat}}=\ve^{(0)}_{\cN}\Omega \pa_v \varphi\delta\varphi$.

To recap, we have found that each sector contributes as
\beq
I'_{\hat f}\Omega^{(0)}&=&- \frac1{8\pi G}\delta \left(\int_{\cN}\ve^{(0)}_{\cN}f C^{(0)} \right)+\int_{\cC} f  \iota_{\pa_v}\theta^{(0)}-\frac1{8\pi G} \delta\left(\int_{\cC}\ve^{(0)}_{\cC}(\Omega\pa_v f- f\pa_v\Omega)\right)\label{C0} \\
I'_{\hat f}\Omega^{(2)}&=&- \frac1{8\pi G}\delta \left(\int_{\cN}\ve^{(0)}_{\cN}fC^{(2)}\right)+\int_{\cC} f  \iota_{\pa_v}\theta^{(2)} \label{C2}\\
I'_{\hat f}\Omega^{\mathsf{mat}}&=&- \frac1{8\pi G}\delta \left(\int_{\cN}\ve^{(0)}_{\cN}f C^{\mathsf{mat}}\right)+\int_{\cC} f  \iota_{\pa_v}\theta^{\mathsf{mat}}.\label{Cm}
\eeq
where we denote $C^{(i)}=(\pa_v V)^2 \tilde C^{(i)}\circ V$, see \eqref{cct}. While the constraint is typically understood as a classical conservation equation for intrinsic data on the null geometry, we see here that it can alternatively be thought of as the vanishing of the sum of stress tensors from each sector. Indeed, the three CFT-like stress tensors
\beq\label{3cs}
C^{(0)}=\pa_v^2\Omega-\mu\pa_v\Omega \qquad C^{(2)}= \Omega \sigma_{a}{}^{b}\sigma_{b}{}^{a} \qquad C^{\mathsf{mat}}=8\pi G\,\Omega T^{\mathsf{mat}}_{vv},
\eeq
satisfy the Raychaudhuri constraint
\beq
C^{(0)}+C^{(2)}+C^{\mathsf{mat}}=0.
\eeq
This interpretation is clear for the matter sector \eqref{Cm}, and for the spin $2$ sector \eqref{C2}, but it can also be applied to the gravitational spin-$0$ sector as shown in \eqref{C0}. We know that the constraint vanishing  corresponds to the gauging of diffeomorphism invariance. What is interesting is that we now understand this gauging as a balancing between the spin-$0$ stress tensor and  the matter like degrees of freedom (matter + spin-$2$). Notice that both the spin-$2$ and matter sector are manifestly positive stress tensors. On the other hand, the spin-$0$ stress tensor is not. 

So we see here that the main feature of the dressing frame is to split completely the charge contributions of the different sectors. The spin-$2$ and the matter sector do not contribute to the charge, but they generate the bulk constraints and the corner flux. On the other hand, the spin-$0$ sector has a multipurpose function. It contributes to the flux at the corner, plus it generates the canonical charge. In an arbitrary time frame, the spin-$0$ sector would appear also in the bulk, and the sectors would be mixed.

By construction, once regrouping all terms together, the final result is
\beq
I'_{\hat f}\Omega^{\mathsf{tot}}=-\delta M_f'+\cF_f,
\eeq
with $\cF_f=\int_{\cC}f \iota_{\pa_v}\theta^{\mathsf{tot}}$ the  flux, $\theta^{\mathsf{tot}}=\theta^{(0)}+\theta^{(2)}+\theta^{\mathsf{mat}}$ and 
\beq\label{primech2}
M_f'=\frac1{8\pi G}\int_{\cN}\ve^{(0)}_{\cN}f C+\frac1{8\pi G} \int_{\cC}\ve^{(0)}_{\cC}(\Omega\pa_v f- f\pa_v\Omega).
\eeq
This is simply the result of Section \ref{S32}. 
The boost charge depends both on a choice of cut $\cC$
and a vector field $f(v,\sigma) \pa_v $ as given by \eqref{primech2}. The function $f$ appearing in the vector field transforms as a scalar in our  phase space. Indeed, consider two vector fields $\xi=f\pa_v$ and $\zeta=h\pa_v$, we have ($\fL_{\hat\lambda_{h}}f=0$)
\beq
&\fL'_{\hat\zeta}\xi=(\fL_{\hat h}+\fL_{\hat\lambda_{h}})\xi=h\pa_vf\pa_v.
\eeq
This proves that $f$ is a scalar on our phase space, and thus $f=\tilde f \circ V$. Using this, we can write the charge in the dressing time as
\beq
M'_{\tilde f}(\cC) \,\hat =\, \frac1{8\pi G} \int_{\cC}\tilde\ve^{(0)}_{\cC} \left(\tilde\Omega \pa_V\tilde f-\tilde f \pa_V \tilde\Omega \right),
\eeq
where we denote $\tilde\ve^{(0)}_{\cC}\circ V=\ve^{(0)}_{\cC}\pa_v V$ the measure in dressing time, and we have now made explicit that this charge depends on the chosen cut $\cC$.

We now specialize this general charge for the boost generator $\tilde f_T(V) =V-T$, where the parameter $T(\sigma)$ is a function on $S$. Calling $M'_{\tilde f_T}(\cC)=M'_{T}(\cC)$, we get
\beq
M'_{T}(\cC) \,\hat =\, \frac1{8\pi G} \int_{\cC}\tilde\ve^{(0)}_{\cC}(\tilde\Omega-(V-T)\pa_V \tilde\Omega)=\int_{\cC}\tilde\ve^{(0)}_{\cC} \tilde{S}_{T},
\eeq
where we defined the boost charge aspect,
\be
\tilde{S}_{T}(V) :=  \frac1{8\pi G} (\tilde\Omega(V)-(V-T)\pa_V \tilde\Omega(V)),
\ee
and we made explicit that it depends on the symmetry parameter $T$, and the time $V$ at which it is evaluated. 
We recall that the choice of a time function $V_0$ is equivalent to specifying the cut $\cC_{V_0}=\{ V= V_0 \}$. So evaluating $\tilde{S}_{T}(V)$ at a specific time $V=V_0$ is the same as evaluating it at the cut $\cC_{V_0}$. We then see from the previous expression that the {\it boost  charge evaluated at the specific cut $\cC= \cC_T$ is the  area in the dressing time}, on-shell
\beq
M'_{T}(\cC_T) \,\hat =\, \frac1{8\pi G} \int_{\cC}\tilde\ve^{(0)}_{\cC}\tilde\Omega.
\eeq
The positivity of this charge is therefore guaranteed by construction.

Since the following discussion is true per point on $\cC$, we focus on the boost charge aspect. Consider then its time derivative
\beq
\pa_V \tilde{S}_{T}(V)\,\hat =\, -\frac1{8\pi G} (V-T)\pa^2_V \tilde\Omega(V) .
\eeq
In the dressing time where $\tilde\mu=0$, we can use the Raychaudhuri constraint \eqref{rmu} to get the flux formula for the boost
\beq\label{pVS}
\pa_V \tilde{S}_{T}(V) \,\hat =\, \frac1{8\pi G} (V-T)\tilde\Omega(V)(\tilde\sigma^2+8\pi G\tilde T^{\mathsf{mat}}_{VV})(V).
\eeq
The key point is that in the dressing time, there is no spin zero contribution to the right-hand side.

Assuming that the boost is evaluated in the future $V\geq T$ of the cut, this leads to two important results: first, one sees that if the radiation is absent and the matter is in its vacuum state, then the flux vanishes and the charge is conserved, even if the null surface is expanding. This means that the dressing time charge is a proper notion of energy that is conserved not only on non-expanding horizons but also in flat space. This solves a longstanding puzzle; see e.g., the recent works \cite{Chandrasekaran:2023vzb, Odak:2023pga}. Second, we see that if we impose the classical null energy condition on the matter fields
 $\tilde T^{\mathsf{mat}}_{VV}\geq 0$, 
we obtain monotonicity of the boost charge aspect
\beq
\pa_V \tilde{S}_{T}(V)\geq 0.
\eeq
The bound is saturated by the shearless configuration, and when matter is in its vacuum state. Notice that this result is intimately tied to the choice of dressing time, where $\tilde\mu=0$. Indeed, it is only in this frame that the Raychaudhuri equation ensures the inequality written above, since in any other frame there would be an additional spin zero contribution to the stress tensor given by $-\mu \pa_v\Omega$, which is  not manifestly positive. The boost charge aspect thus provides a notion of entropy that satisfies positivity and monotonicity even in the presence of expansion.
Consider then the difference between the boost aspects $\tilde S_{T}(V)$ and $\tilde S_{T}(T)$. We find
\be \label{ds}
\Delta \tilde S=\frac1{8\pi G}\left( \Delta \tilde \Omega - (V-T) \tilde\theta(V)\right),
\ee
where we employed the notation $\Delta F= F(V)- F(T)$ for $V\geq T$, to denote the difference of fields evaluated at different cuts $\cC_V$ and $\cC_T$. 

When the second cut $\cC_V$ is expansion free, the difference in the boost charges is the area difference, $\Delta \tilde S =\frac1{8\pi G} \Delta \tilde\Omega$. This is the key mechanism at play 
in the construction of a local entropy formula for Black Hole horizons as given in \cite{Jacobson:2003wv} (see also \cite{Bianchi:2012br}), where the second cut is taken to be at $V=\infty$ and the definition of an event horizon implies that the asymptotic expansion vanishes.
In general, the expansion at the second cut contributes to \eqref{ds} and is subtracted from the area contribution. 

As already remarked, the dressing time coincides with the affine time for marginally trapped surfaces, where the expansion vanishes. An analysis of positivity of the boost charge has been recently performed in \cite{Rignon-Bret:2023fjq}. For arbitrary null hypersurfaces, our result provides a generalization of the boost charge that ensures its positivity and monotonicity. The latter is important for strong subadditivity of entropy, if one can interpret this charge as the modular Hamiltonian of the system. Testing the properties of this charge when  perturbative quantum matter is added, one could also investigate whether this charge is a good candidate to describe the generalized entropy \cite{Wall:2012uf}. We plan to focus on this in the future, and study whether this can be used to generalize the quantum focusing conjecture \cite{Engelhardt:2014gca, Bousso:2015mna, Bousso:2015wca} to arbitrary null hypersurfaces.

\section{Conclusions and Discussion}\label{Conclusions}

In this paper, we have constructed a very general account of the classical dynamics of the geometric data on an arbitrary null geometry, which can be interpreted as a null hypersurface in some classical spacetime. As such, we focus on the presymplectic canonical potential defined on the null geometry, which takes the general form \eqref{thcan}. This presymplectic form  contains the expected spin-2 degrees of freedom consisting of a metric $q$ on the base of the Carroll structure and the shear $\sigma$. In addition, there are spin-0 modes $\Omega$ and $\mu$, as well as spin-1 modes that we have not elaborated upon in this paper.

The spin-2 modes are usefully rewritten in terms of Beltrami differentials. In those terms it is possible to invert the symplectic form to obtain canonical kinematic Poisson brackets on phase space. As detailed in Section \ref{S43}, the result is an intricate pattern that is local on the base and generally extended non-locally in the null fibre  direction. The brackets as given demonstrate that there is non-trivial mixing between the spin-$0$ and spin-$2$ sectors. This mixing is crucial --- in fact we have shown explicitly that were it not for this mixing, the constraint algebra would not be properly represented on phase space. 

In addition to local diffeomorphisms, there is an internal boost symmetry which gives rise to a corner charge, its constraint vanishing identically. We focused in this paper on the diffeomorphisms along the null fibre, and one finds that there exists a combination of them and internal boosts leaving the fibre $\ell$ invariant. The corresponding charge is
\beq
M'_f = \frac1{8\pi G}\int_{\cN}  \ve^{(0)}_{\cN} f C + \frac1{8\pi G}\int_{\cC}\ve^{(0)}_{\cC}\left(\Omega \pa_v f-f\pa_v \Omega\right),
\eeq
where \[C= \pa_v^2\Omega-\mu\pa_v\Omega + \Omega (\sigma_{a}{}^{b}\sigma_{b}{}^a+8\pi G\, T^{\mathsf{mat}}_{vv})\] is the Raychaudhuri constraint. In the above discussion, we showed how to solve the equation $C=0$ non-perturbatively. It is interesting to note that the constraint has contributions from three terms coming from the spin-0, spin-2 and matter sectors, each of which has been established to be a stress tensor density generating null-time reparametrizations. It seems natural to reinterpret the constraint then, not as a conservation law, but as the vanishing of a total stress tensor, expressing the gauging of diffeomorphism invariance along the null geometry. A natural possibility then arises, which is that at the quantum level, this constraint is anomalous in some sectors. This is a future direction worth exploring.

While this is a satisfying result, the Poisson brackets are kinematic and correspondingly the fields involved transform non-trivially under diffeomorphisms. The key to constructing invariant observables is to introduce a dynamical clock variable, which we have shown in Section \ref{S5} to be implicit in the spin-0 sector. A useful way of thinking of the Carroll geometry is that it defines a null congruence through each point of the base, and the dynamical variables simply encode the properties of that congruence. In such terms, it would be a standard assumption that one should use affine time to describe the null congruence. As we showed in Section \ref{S5}, this however is only appropriate in the special case where the expansion vanishes (such as on Killing horizons). In general a better notion  of a dynamical clock is given by the dressing time, in which the canonical variable $\mu$ is set to zero, by a canonical transformation. This is possible precisely because of the presence of the internal boost symmetry which renders $\mu$ as a connection --- locally then, $\mu$ can be rewritten in terms of a dynamical time variable $V$. This addresses precisely the oft-lamented fact that time is pure gauge in gravitational theories. Remarkably, rewriting the theory in terms of this particular clock variable has 
several key properties. First, it is in fact canonically conjugate to the Raychaudhuri constraint on the extended phase space. Second, it renders the other variables (spin-2 gravity modes as well as arbitrary matter fields) invariant under local diffeomorphisms.

The reader may well be confused by this state of affairs. On the one hand we gave a description in which the spin-0 fields were fully dynamical (in the sense of appearing off-shell in the presymplectic form) throughout the bulk of the null geometry. On the other hand, we have an {\it equivalent} description in which the spin-0 degrees of freedom are replaced by a clock and constraint, while other dynamical fields are dressed. In fact, this should be regarded as an expected structure: we see explicitly that solving the constraint, or more precisely reducing to the constraint surface in phase space, corresponds  to requiring the  clock  to be dynamical variables determined by the spin $2$ and matter degrees of freedom while simultaneously dressing all other fields. We have also seen that the spin-0 fields make important contributions to the codimension-2 corner charges. In the language of \cite{Ciambelli:2021vnn,Ciambelli:2021nmv,Freidel:2021dxw,Klinger:2023tgi}, we can interpret the original spin-0 formulation to mean that we are working on an extended phase space, the significance of which the symmetries are represented by Hamiltonian vector fields whose algebra of Hamiltonians closes. From the point of view of the extended phase space, the clock and constraint are dynamical variables, satisfying canonical Poisson brackets. By passing to the second formulation, we have effectively imposed the constraints in the sense that all other fields are dressed. The precise sense in which this has happened though is interesting.  It was not done by setting the constraint to zero, but by performing a canonical transformation in which the constraint becomes a dynamical variable conjugate to the clock. In a quantum theory, the constraint commutes with the dressed operators, but does not commute with the clock. This realizes on a general null geometry the idea that the constraints evolve the system in time, as well known in the timelike ADM formulation of gravity. Thus, interpreting a constraint operator as 'zero' is subtle and irrecovably tied up with what we might want to think of as a notion of time in a diffeomorphism-invariant quantum theory. We remark that to achieve this we had to include the surface tension $\mu$ from the beginning, which is a fundamental ingredient of the theory, that we traded for the dressing time.

When specified to a boost, our new charge $M_{f}'$ has the interesting property of being positive and monotonic, with the inequality saturated by the classical non-radiating configuration. Furthermore, given that the dressing time coincides with the affine time for non-expanding horizons, it reduces to the usual boost charge used to study the laws of thermodynamics on marginally trapped surfaces. In the future, we plan to investigate  whether this charge operator is a good candidate for a generalization of the quantum focusing conjecture to arbitrary null hypersurfaces, and in particular its relationship to the modular Hamiltonian \cite{Wall:2011hj, Bousso:2015wca, Faulkner:2016mzt, Casini:2017roe, Hollands:2019ajl},  given the suggestive form of eq. \eqref{solRaych}.
Finally let us remind the reader that a limitation of our analysis is that it is valid away from creases and caustics. These should be thought of as boundaries of the null regions we are studying here. It would be interesting to understand better how the geometrical matching of null portions through these boundaries is achieved in terms of the corner charges.

As advertised at various points of the manuscript, one of our principle motivations is to study 
the quantization of the system. 
In particular, our results suggest that comparing the various available pathways to quantization may yield important insights and perhaps surprises. 
Another followup project is to extend the work to the Damour constraint, accounting for the transverse diffeomorphisms. We expect that a similar structure will be present, with the Damour constraint conjugate to additional dressing variables playing the proper role of observer devices. 
We expect to report on these topics soon.

\paragraph{Acknowledgements}
This paper is the result of a \emph{focused research group} held at BIRS, Banff, in November 2022, titled "Symmetries of Gravity at the Black Hole Horizon". We are indebted to BIRS for the warm hospitality and for creating such a stimulating environment. We thank Ted Jacobson, Marc Klinger, Jerzy Kowalski-Glikman, and Aron Wall for discussions and feedback. Research at Perimeter Institute is supported in part by the Government of Canada through the Department of Innovation, Science and Economic Development Canada and by the Province of Ontario through the Ministry of Colleges and Universities. LC is grateful to University of Milano Statale (and in particular Antonio Amariti) and University of Trento (and in particular Valter Moretti) for hospitality during the completion of this work. The work of RGL is partially supported by the U.S. Department of Energy under contract DE-SC0015655, and RGL thanks the Perimeter Institute for supporting collaborative visits.

\appendix

\section{Inverse of the Symplectic 2 Form}\label{A1}

To invert the canonical 2 form
\beq
\Omega^{\mathsf{can}}
&=&\frac1{8\pi G}\int_{\cN}\varepsilon_{\cN}^{(0)}
\Big(\delta\Omega\wedge\big(\delta\mu+\bD_{v}\Delta+\Delta_{v}\bD\big)+2D_v(\sqrt{\Omega}\Delta)\wedge\sqrt{\Omega}\bD\Big),
\eeq
we rewrite it as a bi-local expression
\beq
\Omega^{\mathsf{can}}
&=&\frac1{16\pi G} \int_{{\cN}_1}\ve^{(0)}_{{\cN}_1} \int_{{\cN}_2}\ve^{(0)}_{{\cN}_2} \delta Z^\alpha(x_1)\Omega_{\alpha\beta}^{\mathsf{can}}(x_1,x_2)\wedge \delta Z^\beta(x_2)
\eeq
where $Z^\alpha$ is the basis of our fields and $x_1=(v_1,z_1,\bz_1)$. 

Choosing $Z^\alpha(x_1)=( \Omega(x_1), \mu(x_1),\zeta(x_1),\bzeta(x_1))$,
 we get
\beq\label{Om}
\Omega_{\alpha\beta}^{\mathsf{can}}(x_1,x_2)=\frac1{8\pi G}\begin{pmatrix}
0 & 1 & \frac{\bD_{v_1}}{\beta_1} & \frac{\Delta_{v_1}}{\beta_1} \\
-1 & 0 & 0 & 0\\
-\frac{\bD_{v_2}}{\beta_2} & 0 & 0 & -2\frac{\sqrt{\Omega_1\Omega_2}}{\beta_1\beta_2}\overline{D}_{v_1}\\
-\frac{\Delta_{v_2}}{\beta_2} & 0 & -2\frac{\sqrt{\Omega_1\Omega_2}}{\beta_1\beta_2} D_{v_1} & 0
\end{pmatrix}\delta^{(3)}(x_1-x_2)
\eeq
where we have distributed the $x_1$ and $x_2$ dependency in the matrix, and the delta function is defined via
\beq
\int_{{\cN}_1}\ve^{(0)}_{{\cN}_1}\delta^{(3)}(x_1-x_2)f(x_1)=f(x_2).
\eeq

We want to find the distributional inverse. The delicate term is 
\beq
\Omega_{\zeta\bzeta}^{\mathsf{can}}(x_1,x_2)=-\frac1{4\pi G}\frac{\sqrt{\Omega_1\Omega_2}}{\beta_1\beta_2}\overline{D}_{v_1}\delta^{(3)}(x_1-x_2).
\eeq
Its inverse is ($\delta^{(3)}(x_1-x_2)=\delta(v_1-v_2)\delta^{(2)}(z_1-z_2)$)
\beq
(\Omega^{\mathsf{can}}_{\zeta\bzeta}(x_1,x_2))^{-1}=-4\pi G\frac{\beta_1\beta_2}{\sqrt{\Omega_1\Omega_2}}
e^{-2\int_{v_1}^{v_2}\omega_v \rd v}H(v_1-v_2)\delta^{(2)}(z_1-z_2)=-4\pi G \beta_1\beta_2 \bcP_{12},
\eeq
where $H(-v)= -H(v)$ is the odd Heaviside step function satisfying
\beq
\pa_{v_1}H(v_1-v_2)=\delta(v_1-v_2),
\eeq
and we see why the propagator, introduced in \eqref{propa}, is such a key object in this theory.

Using this, we can eventually inverse the matrix \eqref{Om},
\beq
\Omega_{\mathsf{can}}^{\alpha\beta}(x_1,x_2)=4\pi G\begin{pmatrix}
0&-2\delta^{(3)}(x_1-x_2)&0&0
\cr
2\delta^{(3)}(x_1-x_2)& -\left(\Delta_{v_1}\bD_{v_2}\bcP_{12}+\bD_{v_1}\Delta_{v_2}\cP_{12}\right) & \beta_2\Delta_{v_1}\bcP_{12}
&\beta_2\bD_{v_1}\cP_{12}
\cr
0&\beta_1\Delta_{v_2}\cP_{12}
&0&-\beta_1\beta_2\cP_{12}
\cr
0&\beta_1\bD_{v_2}\bcP_{12}
&-\beta_1\beta_2\bcP_{12}&0\end{pmatrix}.
\eeq

With our conventions established in \eqref{conv}, one has $\Omega^{\alpha\beta}_{\mathsf{can}}(x_1,x_2)=-\{Z^\alpha(x_1),Z^\beta(x_2)\}$,
and thus the non-vanishing Poisson brackets are
\beq
\{\Omega_1,\mu_2\}&=&8\pi G\delta^{(3)}(x_1-x_2)\\
\{\mu_1,\mu_2\}&=&4\pi G\left(\Delta_{v_1}\bD_{v_2}\bcP_{12}+\bD_{v_1}\Delta_{v_2}\cP_{12}\right)\\
\{\mu_1,\zeta_2\}&=&-4\pi G\beta_2\Delta_{v_1}\bcP_{12}\\
\{\mu_1,\bzeta_2\}&=&-4\pi G\beta_2\bD_{v_1}\cP_{12}\\
\{\zeta_1,\bzeta_2\}&=&4\pi G\beta_1\beta_2\cP_{12}\\
\{\bzeta_1,\zeta_2\}&=&4\pi G\beta_1\beta_2\bcP_{12}.
\eeq

\section{Composite Brackets}\label{A2}

To explicitly construct brackets among the coframe elements, we exploit the fact that the Poisson bracket is a derivation, and we introduce the pragmatic notation
\beq
\{\zeta(x_1),\bzeta(x_2)\}=\{\zeta_1,\bzeta_2\}=\fL_{\hat\zeta_1}\bzeta_2=I_{\hat\zeta_1}\delta\bzeta_2.
\eeq
We then have, for instance,
\beq
\{\zeta_1,m^2_a\}=\fL_{\hat\zeta_1}m^2_a=I_{\hat \zeta_1}\delta m^2_a=I_{\hat \zeta_1}(\Delta_2\Bm^2_a+\omega_2 m^2_a)=I_{\hat \zeta_1}\Delta_2\Bm^2_a+I_{\hat \zeta_1}\omega_2 m^2_a.
\eeq

Using then
\beq
&I_{\hat\zeta_1}\delta\zeta_2=\{\zeta_1,\zeta_2\}=0\qquad 
I_{\hat\bzeta_1}\delta\bzeta_2=\{\bzeta_1,\bzeta_2\}=0&\\
&I_{\hat\bzeta_1}\delta\zeta_2=4\pi G\beta_1\beta_2 \bcP_{12}\qquad 
I_{\hat\zeta_1}\delta\bzeta_2=4\pi G\beta_1\beta_2 \cP_{12},&
\eeq
we compute
\beq
&I_{\hat\zeta_1}\Delta_2=\frac{I_{\hat\zeta_1}\delta\zeta_2}{\beta_2}=0\qquad I_{\hat\bzeta_1}\bD_2=\frac{I_{\hat\bzeta_1}\delta\bzeta_2}{\beta_2}=0&\\
&I_{\hat\bzeta_1}\Delta_2=\frac{I_{\hat\bzeta_1}\delta\zeta_2}{\beta_2}=4\pi G\beta_1 \bcP_{12}\qquad I_{\hat\zeta_1}\bD_2=\frac{I_{\hat\zeta_1}\delta\bzeta_2}{\beta_2}=4\pi G\beta_1 \cP_{12}&\\
&I_{\hat\zeta_1}\omega_2=\frac{\zeta_2I_{\hat\zeta_1}\delta\bzeta_2-\bzeta_2I_{\hat\zeta_1}\delta\zeta_2}{2\beta_2}=2\pi G\zeta_2\beta_1\cP_{12}\qquad I_{\hat\bzeta_1}\omega_2=\frac{\zeta_2I_{\hat\bzeta_1}\delta\bzeta_2-\bzeta_2I_{\hat\bzeta_1}\delta\zeta_2}{2\beta_2}=-2\pi G\bzeta_2\beta_1\bcP_{12}.&
\eeq

Thus we get
\beq
\{\zeta_1,m^2_a\}=2\pi G\zeta_2\beta_1\cP_{12}m^2_a \qquad \{\zeta_1,m_2^a\}=2\pi G\zeta_2\beta_1\cP_{12}m_2^a,
\eeq
and also
\beq
\{\bzeta_1,m^2_a\}=4\pi G\beta_1\bcP_{12}\Big(\Bm_a^2-\frac{\bzeta_2}{2}m^2_a\Big)\qquad 
\{\bzeta_1,m_2^a\}=-4\pi G\beta_1\bcP_{12}\Big(\Bm^a_2+\frac{\bzeta_2}{2}m_2^a\Big).
\eeq

From these commutators one can derive their conjugates. For instance
\beq
\{\bzeta_1,\Bm^2_a\}=\overline{\{\zeta_1,m^2_a\}}=2\pi G\bzeta_2\beta_1\bcP_{12}\Bm^2_a.
\eeq
Then we have
\beq
\fL_{m_a^1}\zeta_2=\{m_a^1,\zeta_2\}=-\{\zeta_2,m_a^1\}=-2\pi G\zeta_1\beta_2\cP_{21}m^1_a=2\pi G\zeta_1\beta_2\bcP_{12}m^1_a,
\eeq
and so on.

Repeating the same steps as before for $I_{\hat m_a^1}$ etc., we arrive to the coframe commutators. We display the two fundamental blocks, while the others can be derived by conjugation
\beq
\{m^1_a,m_2^b\}&=&
\pi G\left(-2\zeta_1\bcP_{12}m^1_a\Bm_2^b+2\zeta_2
\cP_{12}\Bm_a^1 m_2^b-\bzeta_1\zeta_2\cP_{12}m^1_a m_2^b-\zeta_1\bzeta_2\bcP_{12}m^1_a m_2^b\right)\\
\{\Bm^1_a,m_2^b\}&=&\pi G\left(-4\bcP_{12}m_a^1\Bm_2^b+2\zeta_1\bcP_{12}\Bm^1_a\Bm_2^b+\bzeta_1\zeta_2\cP_{12}\Bm^1_a m_2^b-2\bzeta_2\bcP_{12}m_a^1 m_2^b+\zeta_1\bzeta_2\bcP_{12}\Bm^1_a m_2^b\right).\nonumber
\eeq

From here, a lengthy yet straightforward computation leads to the commutators
\beq
\{m^1_a\Bm_1^b,m^2_c \Bm_2^d\}&=&-4\pi G\cP_{12}\Bm_a^1\Bm_1^b m^2_c m_2^d-4\pi G\bcP_{12}m^1_a m_1^b \Bm_c^2\Bm_2^d,\\
\{\Bm^1_a m_1^b,m^2_c \Bm_2^d\}&=&4\pi G\bcP_{12}m_a^1 m_1^b\Bm^2_c\Bm_2^d+4\pi G\cP_{12}\Bm^1_a \Bm^b_1 m^2_c m_2^d.
\eeq

Given our parameterization, the projector $\bq_{cb}\bq^{ba}=q_c{}^a$ is given by $Diag(0,1,1)$, and thus a good consistency check is that one obtains
\beq
\{q^{1 \, b}_{a},q^{2 \, d}_{c}\}=0\qquad \{\epsilon^{1 \, b}_{a},q^{2 \, d}_{c}\}=0.
\eeq
On the other hand, one has
\beq
\{\bq^1_{ab},\bq^2_{cd}\}=16\pi G\bcP_{12}m^1_a m_b^1\Bm^2_c\Bm^2_d+16\pi G\cP_{12}\Bm_a^1 \Bm^1_b m^2_c m^2_d,
\eeq
proving that our degenerate metric satisfies a non-commutative geometric structure. This is further confirmed by
\beq
\{\epsilon^{1 \, b}_{a},\epsilon^{2 \, d}_{c}\}=-16\pi G\cP_{12}\Bm_a^1\Bm_1^b m^2_c m_2^d-16\pi G\bcP_{12}m^1_a m^b_1\Bm^2_c\Bm_2^d,
\eeq
which is the result reported in the main text.

\section{Shear Bracket}\label{A3}

We here show the bracket of the shear tensor, as given in \eqref{qdq}.

A brute force computation yields 
\beq
\{m^1_a,\Delta_{v_2}\}&=&2\pi G\left(\zeta_1\pa_{v_2}\bcP_{12} m^1_a-\zeta_1\zeta_2\bD_{v_2}\bcP_{12} m^1_a+2\zeta_2 \Delta_{v_2}\cP_{12} \Bm^1_a-\bzeta_1\zeta_2\Delta_{v_2}\cP_{12} m^1_a\right)\\
\{m^1_a,\bD_{v_2}\}&=&2\pi G\left(2\pa_{v_2}\cP_{12}\Bm^1_a-\bzeta_1\pa_{v_2}\cP_{12} m^1_a-2\bzeta_2 \Delta_{v_2}\cP_{12}\Bm^1_a+\bzeta_1\bzeta_2\Delta_{v_2}\cP_{12}m^1_a+\zeta_1\bzeta_2\bD_{v_2}\bcP_{12}m^1_a\right),\nonumber
\eeq
and
\beq
\{\Delta_{v_1},\Delta_{v_2}\}&=&4\pi G\left(-\zeta_1\zeta_2\Delta_{v_1}\bD_{v_2}\bcP_{12}-\zeta_1\zeta_2\bD_{v_1}\Delta_{v_2}\cP_{12}+\zeta_2\Delta_{v_2}\pa_{v_1}\cP_{12}+\zeta_1\Delta_{v_1}\pa_{v_2}\bcP_{12}\right)\\
\{\bD_{v_1},\Delta_{v_2}\}&=&4\pi G\left(\pa_{v_1}\pa_{v_2}\bcP_{12}-\zeta_2\bD_{v_2}\pa_{v_1}\bcP_{12}-\bzeta_1\Delta_{v_1}\pa_{v_2}\bcP_{12}+\bzeta_1\zeta_2\bD_{v_1}\Delta_{v_2}\cP_{12}+\bzeta_1\zeta_2\Delta_{v_1}\bD_{v_2}\bcP_{12}\right).\nonumber
\eeq

We have to evaluate
\beq
\{\sigma^{1b}_a,\sigma^{2d}_c\}&=&\{\Delta_{v_1} \Bm^1_a \Bm_1^b+  \bar\Delta_{v_1} m^1_a m_1^b,\Delta_{v_2} \Bm^2_c \Bm_2^d+  \bar\Delta_{v_2} m^2_c m_2^d\}\cr
&=&\{\Delta_{v_1} \Bm^1_a \Bm_1^b,\Delta_{v_2} \Bm^2_c \Bm_2^d\}+\{ \bar\Delta_{v_1} m^1_a m_1^b,\Delta_{v_2} \Bm^2_c \Bm_2^d\}\\
&&+\{\Delta_{v_1} \Bm^1_a \Bm_1^b,\bar\Delta_{v_2} m^2_c m_2^d\}+\{ \bar\Delta_{v_1} m^1_a m_1^b, \bar\Delta_{v_2} m^2_c m_2^d\}\cr
&=&\Delta_{v_2}\{\Delta_{v_1} \Bm^1_a \Bm_1^b,\Bm^2_c \Bm_2^d\}+\{\Delta_{v_1} \Bm^1_a \Bm_1^b,\Delta_{v_2}\} \Bm^2_c \Bm_2^d+\Delta_{v_2}\{ \bar\Delta_{v_1} m^1_a m_1^b, \Bm^2_c \Bm_2^d\}\cr
&&+\{ \bar\Delta_{v_1} m^1_a m_1^b,\Delta_{v_2}\} \Bm^2_c \Bm_2^d+\bar\Delta_{v_2}\{\Delta_{v_1} \Bm^1_a \Bm_1^b, m^2_c m_2^d\}+\{\Delta_{v_1} \Bm^1_a \Bm_1^b,\bar\Delta_{v_2} \}m^2_c m_2^d\cr
&&+\bar\Delta_{v_2}\{ \bar\Delta_{v_1} m^1_a m_1^b,  m^2_c m_2^d\}+\{ \bar\Delta_{v_1} m^1_a m_1^b, \bar\Delta_{v_2}\} m^2_c m_2^d.
\eeq
Putting together all these brackets, there are many cancellations, and the remaining non-vanishing contributions are
\beq
\{\sigma^{1b}_a,\sigma^{2d}_c\}
&=&4\pi G\Big(\Bm^1_a\Bm_1^b \epsilon^{2d}_c\Big(\Delta_{v_1}\Delta_{v_2}\bzeta_1\cP_{12}+\Delta_{v_2}\pa_{v_1}\cP_{12}-\Delta_{v_2}\zeta_1 \bD_{v_1}\cP_{12}\Big)\cr
&&+\epsilon^{1b}_a \Bm^2_c \Bm_2^d\Big(\Delta_{v_1}\Delta_{v_2}\bzeta_2
\bcP_{12}+\Delta_{v_1}\pa_{v_2}\bcP_{12}-\Delta_{v_1}\zeta_2 \bD_{v_2}\bcP_{12}\Big)-\bD_{v_1}\Delta_{v_2}\cP_{12}\epsilon^{1b}_a\epsilon^{2d}_c\cr
&&+m^1_a m_1^b\Bm^2_c \Bm_2^d\Big(\Delta_{v_2}\bzeta_2\pa_{v_1}\bcP_{12} -\Delta_{v_2}\bzeta_2\bzeta_1\Delta_{v_1}\bcP_{12}+\pa_{v_1}\pa_{v_2}\bcP_{12}\cr
&&-\zeta_2\bD_{v_2}\pa_{v_1}\bcP_{12}-\bzeta_1\Delta_{v_1}\pa_{v_2}\bcP_{12}+\bzeta_1\zeta_2\Delta_{v_1}\bD_{v_2}\bcP_{12}\cr
&&+\bD_{v_1}\zeta_1\pa_{v_2}\bcP_{12}-\bD_{v_1}\zeta_1\zeta_2\bD_{v_2}\bcP_{12}+\bD_{v_1}\Delta_{v_2}\zeta_1\bzeta_2\bcP_{12}\Big)+c.c.\Big),
\eeq
where $c.c.$ denotes complex conjugation.\footnote{The tensor $\epsilon_a{}^b$ is purely imaginary $\overline{\epsilon}_a{}^b=-\epsilon_a{}^b$.} Using the relation $2\omega_{v}=\zeta\bD_{v}-\bzeta\Delta_{v}$, and recalling $D_{v}=\pa_v-2\omega_v$ and $\oD_{v}=\pa_v+2\omega_v$, we obtain
\beq
\{\sigma^{1b}_a,\sigma^{2d}_c\}
&=&4\pi G\Big(\Bm^1_a\Bm_1^b \epsilon^{2d}_c\Delta_{v_2}D_{v_1}\cP_{12}+\epsilon^{1b}_a \Bm^2_c \Bm_2^d\Delta_{v_1}D_{v_2}\bcP_{12}-\bD_{v_1}\Delta_{v_2}\cP_{12}\epsilon^{1b}_a\epsilon^{2d}_c\cr
&&+m^1_a m_1^b\Bm^2_c \Bm_2^d\overline{D}_{v_1}D_{v_2}\bcP_{12}+C.C.\Big).
\eeq

\section{Dressing Frame Symplectic Structure}\label{appendixclocksymp}

In this appendix we work out the symplectic form in the dressing time, to show the detailed computation used in section \ref{S52}. An important property used in the derivation below is the fact that $\delta$ commutes with the coordinate time derivative, $\delta\pa_v=\pa_v\delta$, and from this one can establish that \beq
([\pa_V,\delta] \tilde O)\circ V=0.
\eeq

Let us begin with the matter field, whose presymplectic potential has been introduced in \eqref{omma}.
Using \eqref{phi}, we get
\beqn
\Omega^{\mathsf{mat}}
&=&\int_{\cN}\ve^{(0)}_{\cN}\delta(\Omega\pa_v \varphi)\wedge \delta\varphi\cr
&=&\int_{\cN}\ve^{(0)}_{\cN}(\delta\Omega\pa_v \varphi+\Omega\pa_v\delta\varphi)\wedge (\delta \tilde\varphi+ \pa_V \tilde\varphi \delta V)\circ V\cr
&=&\int_{\cN}\ve^{(0)}_{\cN}((\delta\tilde\Omega+\pa_V \tilde\Omega \delta V)\circ V\pa_v \varphi+\tilde\Omega\pa_v((\delta\tilde\varphi+\pa_V\tilde\varphi \delta V)\circ V))\wedge (\delta \tilde\varphi+ \pa_V \tilde\varphi \delta V)\circ V\nonumber\\
&=&\int_{\cN}\ve^{(0)}_{\cN}\left(\pa_v V \delta(\tilde\Omega\pa_V\tilde\varphi)\wedge\delta\tilde\varphi\circ V+\delta(\pa_v V \tilde\Omega(\pa_V\tilde\varphi)^2)\circ V \wedge \delta V+\pa_v (\delta V\wedge(\tilde\Omega\pa_V \tilde\varphi\delta\tilde\varphi)\circ V)\right).
\nonumber\eeqn
This is the matter contribution: the first term appears in the first line of \eqref{bulkom}, and gives rise to $\int_{\cN}\pa_v V\delta\tilde\theta^{\mathsf{mat}}\circ V$, the second term is the matter contribution to the constraint on the second line of \eqref{bulkom}, $\frac1{8\pi G}\int_{\cN}\ve^{(0)}_{\cN}\delta(\pa_v V \tilde C^{\mathsf{mat}}\circ V)\wedge \delta V$, and finally the third term is the corner contribution appearing in \eqref{Cornersymp}, which becomes \eqref{omatc}.

Next, we consider the spin-2 contribution to \eqref{2fo}, which is processed in a very similar way to the matter sector ($q$ is like $\varphi$ while $\sigma$ is like $\pa_v \varphi$) to give\footnote{We have used the identity $\tilde\sigma^{ab}\pa_V\tilde q_{ab}=2\tilde\sigma^2$.}
\beqn
\Omega^{\mathsf{can}(2)}
&=&\frac1{16\pi G}\int_{\cN}\ve^{(0)}_{\cN}\delta(\Omega\sigma^{ab})\wedge \delta q_{ab}\\
&=&\frac1{16\pi G}\int_{\cN}\ve^{(0)}_{\cN}\big(\pa_v V \delta(\tilde\Omega\tilde\sigma^{ab})\wedge\delta\tilde q_{ab}\circ V+2\delta(\pa_v V \tilde\Omega\tilde\sigma^2)\circ V \wedge \delta V\cr
&&+\pa_v (\delta V\wedge(\tilde\Omega \tilde\sigma^{ab}\delta\tilde q_{ab})\circ V)\big).
\eeqn
This displays exactly the same pattern as the matter contribution: the first term appears in the first line of \eqref{bulkom}, and gives rise to $\int_{\cN}\pa_v V\delta\tilde\theta^{(2)}\circ V$, the second term is the spin-$2$ contribution to the constraint on the second line of \eqref{bulkom}, $\frac1{8\pi G}\int_{\cN}\ve^{(0)}_{\cN}\delta(\pa_v V \tilde C^{(2)}\circ V)\wedge \delta V$, and finally the third term is the corner contribution appearing in \eqref{Cornersymp}, which becomes \eqref{os2c}.

The spin-$0$ part is different from the rest. Starting again from \eqref{2fo}, and using \eqref{dmu}, we get
\beq
\Omega^{\mathsf{can}(0)}
&=&-\frac1{8\pi G}\int_{\cN}\ve^{(0)}_{\cN}\delta\mu\wedge\delta\Omega\\
&=&-\frac1{8\pi G}\int_{\cN}\ve^{(0)}_{\cN}\pa_v\left(\frac{\pa_v\delta V}{\pa_v V}\right)\wedge \delta\tilde\Omega\circ V-\frac1{8\pi G}\int_{\cN}\ve^{(0)}_{\cN}\pa_v\left(\frac{\pa_v\delta V}{\pa_v V}\wedge \pa_V\tilde\Omega\circ V \delta V\right)\nonumber\\
&&+\frac1{8\pi G}\int_{\cN}\ve^{(0)}_{\cN} \pa_V^2\tilde\Omega \circ V\delta\pa_v V\wedge \delta V\nonumber\\
&=&\frac1{8\pi G}\int_{\cN}\ve^{(0)}_{\cN}\delta\left(\pa_V^2\tilde\Omega\circ V\pa_v V\right)\wedge \delta V-\frac1{8\pi G}\int_{\cN}\ve^{(0)}_{\cN}\pa_v\left(\pa_V\delta\tilde\Omega\circ V\wedge \delta V\right)\nonumber\\
&&-\frac1{8\pi G}\int_{\cN}\ve^{(0)}_{\cN}\pa_v\left(\frac{\pa_v\delta V}{\pa_v V}\wedge(\delta\tilde\Omega+\pa_V\tilde\Omega\delta V)\circ V\right)\nonumber\\
&=&\frac1{8\pi G}\int_{\cN}\ve^{(0)}_{\cN}\delta\left(\pa_V^2\tilde\Omega\circ V\pa_v V\right)\wedge \delta V-\frac1{8\pi G}\int_{\cN}\ve^{(0)}_{\cN}\pa_v\left(\frac{\pa_v\delta V}{\pa_v V}\wedge\delta\Omega-\delta V\wedge \delta\left(\frac{\pa_v\Omega}{\pa_v V}\right)\right).\nonumber
\eeq
We thus see that the first term  is the spin-$0$ contribution to the constraint on the second line of \eqref{bulkom}, $\frac1{8\pi G}\int_{\cN}\ve^{(0)}_{\cN}\delta(\pa_v V \tilde C^{(0)}\circ V)\wedge \delta V$, while the last term is the corner contribution \eqref{oc0}.

There are extra contributions to the spin-$0$ part when considering a generic clock, since one has
\beq
\delta\mu=\pa_v\delta V\tilde\mu\circ V+\pa_V \tilde\mu \circ V\pa_v V\delta V+\pa_v V\delta\tilde\mu\circ V+\pa_v\left(\frac{\pa_v\delta V}{\pa_v V}\right).
\eeq
There are three additional terms, so that the final result is
\beq
\Omega^{\mathsf{can}(0)}
&=&\frac1{8\pi G}\int_{\cN}\ve^{(0)}_{\cN}\delta\left((\pa_V^2\tilde\Omega-\tilde\mu \pa_V\tilde\Omega)\circ V\pa_v V\right)\wedge \delta V\\
&&-\frac1{8\pi G}\int_{\cN}\ve^{(0)}_{\cN}\pa_v V\delta\tilde\mu\wedge \delta\tilde\Omega\circ V\\
&&-\frac1{8\pi G}\int_{\cN}\ve^{(0)}_{\cN}\pa_v\left(\delta V\wedge \tilde\mu \delta\tilde\Omega\circ V\right)\\
&&-\frac1{8\pi G}\int_{\cN}\ve^{(0)}_{\cN}\pa_v\left(\frac{\pa_v\delta V}{\pa_v V}\wedge\delta\Omega-\delta V\wedge \delta\left(\frac{\pa_v\Omega}{\pa_v V}\right)\right).
\eeq
We see that in a generic time frame we restore the bulk $\tilde\mu$-contributions to the constraint (first line), to the bulk presymplectic structure (second line), and to the corner term (third line). 

\bibliographystyle{uiuchept}
\bibliography{ArxivV3.bib}

\end{document}